\documentstyle[aps,pre,epsf,epsfig,rotate,cite,multicol]{revtex}

\begin{document}

\title{Effect of nonstationarities on detrended fluctuation analysis}

\author{Zhi~Chen$^1$, Plamen~Ch.~Ivanov$^{1, 2}$, Kun~Hu$^1$, H.~Eugene~Stanley$^1$}

\address{ $^1$ Center for Polymer Studies and Department of Physics,
               Boston University, Boston, Massachusetts 02215
\\$^2$ Harvard Medical School, Beth Israel Deaconess Medical Center, Boston,
               Massachusetts 02215\\}

\maketitle

\begin{abstract}

Detrended fluctuation analysis (DFA) is a scaling analysis method used to 
quantify long-range power-law correlations in signals. 
Many physical and biological signals are ``noisy'', heterogeneous and  
exhibit  different types of nonstationarities, which can affect the 
correlation properties of these signals.  We systematically
study the effects of three  types of nonstationarities often encountered
in real data. Specifically, we consider nonstationary sequences formed in
three ways: (i) 
stitching together segments of data obtained from
discontinuous experimental recordings, or removing some noisy and unreliable
 parts from continuous recordings and stitching together the remaining  parts
--- a ``cutting'' procedure commonly used in preparing data prior to signal
analysis; (ii) adding to a signal with known correlations a tunable
concentration of random outliers or spikes with different amplitude, and (iii)
generating  a signal
comprised of segments with different properties --- e.g. different standard
deviations or different correlation exponents. We 
compare the difference between the scaling results obtained for stationary
correlated signals and correlated signals with these
three types of nonstationarities. We find that introducing
nonstationarities to stationary correlated signals leads to the appearance of
crossovers in the scaling behavior and we study how the characteristics of
these crossovers depend on: (a) the fraction and size of the parts cut
out from the signal; (b) the concentration  of spikes and their 
amplitudes;
(c) the proportion between segments with different standard deviations or
different correlations; and (d) the correlation properties of the stationary
 signal. We show how to develop strategies for pre-processing ``raw'' data 
 prior to analysis, which will minimize the effects of nonstationarities on the
 scaling properties of the data and how to interpret the results
 of DFA for complex signals with different local characteristics.

\end{abstract}
\begin{multicols}{2}
\section{Introduction} \label{secintr}

In recent years, there has been growing evidence indicating 
 that many physical and biological
systems have no characteristic length scale and exhibit long-range power-law
correlations. Traditional approaches such as the power-spectrum and
 correlation analysis are suited to quantify correlations in stationary
 signals \cite{non,witt1994}. However, many signals which are outputs of complex 
 physical and biological systems are nonstationary --- the mean,
  standard deviation and higher moments, or the correlation functions are
 not invariant under time translation \cite{non,witt1994}. Nonstationarity, 
 an important aspect of complex variability, can often be associated
 with different trends in the signal or heterogeneous segments (patches)
 with different local statistical properties. To address this problem, 
 detrended fluctuation analysis (DFA) was developed to accurately quantify
 long-range power-law correlations embedded in a nonstationary time series
 \cite{CKDFA1,taqqu95}. This method provides a    
 single quantitative parameter --- the scaling exponent $\alpha$ --- to
 quantify the correlation properties of 
a signal. One advantage of the DFA method is that it allows the detection of
 long-range power-law correlations in noisy signals with embedded polynomial
 trends that can mask the true correlations in the fluctuations of a
 signal. The DFA method has been successfully applied to research fields
 such as
 DNA\cite{CKDFA1,cknature1992,rmsCK,SVDFA1,SMDFA1,genenuovodna1994,mantegnaprl1994,CKfractal,ckdnapha1995,solomdnafractal1995,mantegnaprl1996,Buldyrev,HE1},
 cardiac dynamics
\cite{crossoverCK,iyengaramjphsiolreg,plamennature1996,HOcirc1997,plamenphsa1998,barbiheartchaos1998,plamenuropl1999,Pikkujamsaheartcir1999,plamennature1999,solomrev1999,Genephsa1999,ashkenazyheartfrac1999,makikallioheartamjcardiol1999,Absil1999,solomphsa1999,toweillheartmed2000,bundesleep2000,Laitio2000,yose2000,Yosef2001,plamenchaos2001},
human gait \cite{hos},
 meteorology \cite{Ivanovameteo1999_12}, climate
 temperature fluctuations \cite{Bundeatm,bundetem,talknertem2000}, river flow
 and discharge \cite{Montanari2000,Matsoukas2000}, neural
 receptors in biological systems \cite{bahareuph2001}, and economics
 \cite{Liu97,vandewallephsa1997,vandewallepre1998,Liu99,janosiecopha1999,ausloosphsa1999_12,robertoecopha1999,Vandewalle1999,grau-carles2000,ausloosphsa2000_9,ausloosphsa2000_10,ausloospre2001,ausloosIntJModPhys2001}.
 The DFA method may also help identify different states of 
the same system with different scaling behavior ---
e.g., the scaling exponent $\alpha$ for heart-beat intervals is different for
 healthy and sick
 individuals \cite{crossoverCK,ashkenazyheartfrac1999} as
 well as for waking and sleeping states \cite{plamenuropl1999,bundesleep2000}.

To understand the intrinsic dynamics of a given system, it is important to
analyze and correctly interpret its output signals. One of the common challenges is
that the scaling exponent is not always constant (independent of scale) and
crossovers often exist --- i.e., the value of the scaling exponent $\alpha$
differs 
for different ranges of scales
\cite{crossoverCK,iyengaramjphsiolreg,plamenuropl1999,plameneuph1998,janphysica2001}.
A crossover is usually due to a change in the correlation
properties of the signal at different time or space scales, though it 
 can also be a result of nonstationarities in the
signal. A recent work considered different types of
nonstationarities associated with different trends (e.g., polynomial,
sinusoidal and power-law trends) and systematically studied their
effect on the scaling behavior of long-range correlated signals
 \cite{kun}. Here we consider 
the effects of three other types of nonstationarities which are often
encountered in real data or result from  ``standard'' data pre-processing
approaches.  
\\
\\
({\it i}) {\it Signals with segments removed}\\
First we consider a type of nonstationarity caused by 
 discontinuities in signals. Discontinuities may arise from the nature of 
experimental recordings -- e.g., stock exchange data are not recorded during
the nights, weekends and holidays \cite{Liu97,vandewallephsa1997,vandewallepre1998,Liu99,janosiecopha1999,ausloosphsa1999_12,robertoecopha1999,Vandewalle1999}. Alternatively,
discontinuities may be caused by
the fact that some noisy and unreliable portions of continuous recordings
must be discarded, as
 often occurs when analyzing physiological signals \cite{crossoverCK,iyengaramjphsiolreg,plamennature1996,HOcirc1997,plamenphsa1998,barbiheartchaos1998,plamenuropl1999,Pikkujamsaheartcir1999,plamennature1999,solomrev1999,Genephsa1999,ashkenazyheartfrac1999,makikallioheartamjcardiol1999,Absil1999,solomphsa1999,toweillheartmed2000,bundesleep2000,Laitio2000,yose2000,Yosef2001,plamenchaos2001}.
In this case, a common pre--processing procedure is to cut out the noisy,  
unreliable parts of the recording and stitch together the remaining 
informative segments before any statistical analysis is performed. One 
immediate problem is how such cutting procedure will affect the scaling 
properties of long-range correlated signals. A careful consideration should 
be given when 
interpreting results obtained from scaling analysis, so that an
accurate estimate of the true correlation properties of the original signal
may be obtained.
\\       
\\
({\it ii}) {\it Signals with random spikes}\\
A second type of nonstationarity is due to the existence of spikes in data,
which is very common in real life signals
\cite{crossoverCK,iyengaramjphsiolreg,plamennature1996,HOcirc1997,plamenphsa1998,barbiheartchaos1998,plamenuropl1999,Pikkujamsaheartcir1999,plamennature1999,solomrev1999,Genephsa1999,ashkenazyheartfrac1999,makikallioheartamjcardiol1999,Absil1999,solomphsa1999,toweillheartmed2000,bundesleep2000,Laitio2000,yose2000,Yosef2001,plamenchaos2001,hos}. 
Spikes may arise from  external conditions which have little to do with the
intrinsic dynamics of the system. In this case,   
we must distinguish the spikes from normal intrinsic
fluctuations in the system's output and filter them out 
when we attempt to quantify correlations. Alternatively, spikes may
arise from the intrinsic dynamics of the system, rather than being an
epiphenomenon of external conditions. In this second case,
careful considerations should be given as to whether the spikes should be
filtered out when estimating correlations in the signal, since such
``intrinsic'' spikes may be related to the properties of the noisy 
fluctuations. Here, we consider only the simpler case -- namely, when the
spikes are independent of the fluctuations in the signal. The problem is how
spikes affect the scaling behavior of correlated signals, e.g., what kind of 
crossovers they may possibly cause. We also demonstrate  to what extent
features of the crossovers depend on the statistical properties of the
spikes. Furthermore, we show how to recognize if a crossover indeed indicates 
a transition from one type of underlying correlations to a different type, or 
if the crossover is due to spikes without any transition in the dynamical
properties of the fluctuations. 
\\
\\
({\it iii}) {\it Signals with different local behavior}\\
A third type of nonstationarity is associated with the presence of segments in a
signal which exhibit different local statistical properties, e.g., different
local standard deviations or different local correlations. Some examples
include: (a) 24 hour records of heart rate fluctuations are characterized by
segments with larger standard deviation during stress and physical activity 
and segments with smaller standard deviation during rest
\cite{plamennature1996}; (b) studies of DNA 
show that coding and non-coding regions are characterized by different types 
of correlations \cite{cknature1992,SMDFA1}; (c) brain wave analysis of different sleep
stages (rapid eye movement [REM] sleep, light sleep and deep sleep) indicates
that the signal during each stage may have different correlation 
properties \cite{SLEEP1}; (d) heartbeat signals during different sleep stages
 exhibit different scaling properties\cite{bundesleep2000}. For such complex
signals, results  from scaling analysis often reveal a very
complicated structure. It is a challenge to quantify the correlation
properties of these signals. Here, we take a first step toward understanding the
scaling behavior of such signals.

We study these three types of nonstationarities embedded in correlated 
signals. We apply the DFA method to stationary correlated signals and  
identical signals with
 artificially imposed nonstationarities, and compare the difference
 in the scaling results. ({\it i}) We find that cutting segments from a signal and
 stitching together the remaining parts does not affect the scaling for
 positively correlated signals. However, this cutting procedure
 strongly affects anti-correlated signals, leading to a crossover from an
 anti-correlated regime (at small scales) to an uncorrelated regime (at large 
scales). ({\it ii}) For the correlated signals with superposed random spikes, we
 find that the scaling behavior is a superposition of the scaling of the
 signal and the apparent scaling of the spikes. We analytically
 prove this superposition relation by introducing a {\it superposition rule\/}. 
 ({\it iii}) For the case of complex signals comprised of segments with
 different local properties, we find that their scaling behavior is a
 superposition of the scaling of the different components --- 
each component containing only the segments exhibiting identical statistical
 properties. Thus, to obtain the scaling properties of the signal, we 
 need only to examine the properties 
 of each component --- a much simpler task than analyzing the
 original signal.

The layout of the paper is as follows: In Sec.~\ref{secpuren}, 
we describe how we generate signals with desired long-range correlation 
properties and introduce the DFA method to quantify these correlations. In
Sec.~\ref{seccut}, we compare the scaling properties of correlated signals 
before and after removing some segments from the signals. In
Sec.~\ref{secrspikes}, we consider the effect of random spikes on correlated
signals. We show that the superposition of spikes and signals can be explained
 by a superposition rule derived in Appendix \ref{secadd}. In 
Sec.~\ref{secmix}, we study signals comprised of segments with different 
local behavior. We systematically examine all resulting crossovers, their 
conditions of existence, and their typical characteristics associated with the
different types of nonstationarity. We summarize our
findings in Sec.~\ref{Conclusion}. 

\section{Method}\label{secpuren}

Using a modified Fourier filtering method\cite{MFFM}, we generate stationary
uncorrelated, correlated, and anti-correlated signals $u(i)$
($i=1,2,3,...,N_{\mbox{\scriptsize max}}$) with a standard 
deviation $\sigma=1$. This method consists of the following steps:

(a) First, we generate an uncorrelated and Gaussian distributed sequence
{$\eta(i)$} and calculate the Fourier transform coefficients {$\eta(q)$}.

(b) The desired signal $u(i)$ must exhibit correlations, which are
defined by the form of the power spectrum
\begin{equation}
S(q)=\langle u(q)u(-q) \rangle \sim q^{-(1-\gamma)},
\label{M1}
\end{equation}    
where {$u(q$)} are the Fourier transform coefficients of {$u(i)$} and $\gamma$
is the correlation exponent. Thus, we generate {$u(q$)} using the following
transformation: 
\begin{equation}
u(q)=[S(q)]^{1/2}\eta(q),
\label{M2}
\end{equation}
where $S(q)$ is the desired power spectrum in Eq.~(\ref{M1}).

(c) We calculate the inverse Fourier transform of {$u(q$)} to obtain
$u(i)$. 

We use the stationary correlated signal $u(i)$ to generate signals with
different types of nonstationarity and apply the DFA method\cite{CKDFA1} to quantify
correlations in these nonstationary signals. 

Next, we briefly introduce the DFA method, which involves the following
steps\cite{CKDFA1}:  

({\it i})  Starting with a correlated signal  
$u(i)$, where $i=1,..,N_{max}$ and $N_{max}$ is the length of the signal, 
we first integrate the signal $u(i)$ and obtain
$y(k)\equiv\sum_{i=1}^{k}[u(i)-\langle u \rangle]$, where $\langle u
\rangle$ is the mean. 
 
({\it ii}) The integrated signal $y(k)$ is divided into boxes of equal
length $n$.
 
\begin{figure}
\vspace*{0.truein}
\centerline{
\epsfysize=6.5cm\epsfbox{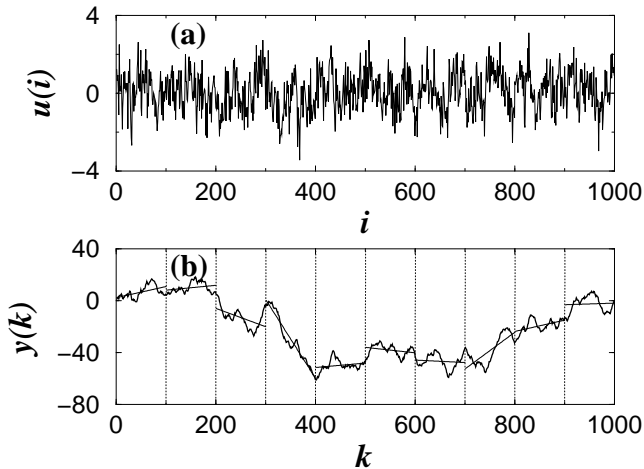}}
\vspace*{0.5cm}
\caption{(a) The correlated signal $u(i)$.
(b) The integrated signal: $y(k)=\sum_{i=1}^k[u(i)-\langle u
\rangle]$. The vertical dotted lines indicate a box of size
        $n=100$, the solid straight lines segments are the estimated
         linear ``trend'' in each box by least-squares fit.}
\label{f.dfa-box}
\end{figure}

({\it iii}) In each box of length $n$, we
fit $y(k)$, using a polynomial function of order $\ell$ which represents the 
{\it trend\/} in that box. The $y$
coordinate of the fit line in each box is denoted by $y_n(k)$ (see
Fig.~\ref{f.dfa-box}, where linear fit is used). Since we use a polynomial
fit of order $\ell$, we  
denote the algorithm as DFA-$\ell$.

({\it iv}) The integrated signal $y(k)$ is detrended by
subtracting the local trend $y_n(k)$ in each box of length $n$.
 
({\it v}) For a given box size $n$, the root mean-square (r.m.s.) 
fluctuation for this integrated and detrended signal is
calculated:
\begin{equation}
 F(n)\equiv\sqrt{{1\over {N_{max}}}\sum_{k=1}^{N_{max}}[y(k)-y_n(k)]^2}.
\label{F2}
\end{equation} 
({\it vi}) The above computation is repeated for a broad range of scales 
(box sizes $n$) to provide a relationship between $F(n)$ and the box
size $n$.

A power-law relation between the average root-mean-square
fluctuation function $F(n)$ and the box size $n$ indicates
the presence of scaling: $F(n) \sim n^{\alpha}$. The fluctuations can be
characterized by a scaling exponent $\alpha$, a self-similarity
parameter which represents the long-range power-law
correlation properties of the signal. If $\alpha=0.5$, there is no correlation
and the signal is  uncorrelated  (white noise); if
$\alpha < 0.5$, the signal is anti-correlated; if
$\alpha >0.5$, the signal is correlated\cite{gamma}. 

We note that for anti-correlated signals, the scaling exponent obtained
from the DFA method overestimates the true correlations at small scales\cite{kun}. 
To avoid this problem, one needs first to integrate the original anti-correlated
signal and then apply the DFA method\cite{kun}. The correct scaling 
exponent can thus be obtained from the relation between $n$ and $F(n)/n$
[instead of $F(n)$]. 
 In the following sections, we first integrate the signals under
 consideration, then apply DFA-2 to remove linear trends in these 
integrated signals. In order to provide a more accurate estimate of $F(n)$, 
the largest box size $n$ we use is $N_{max}/10$, where $N_{max}$ is the
total number of points in the signal.    

We compare the results of the DFA method obtained from the nonstationary
signals with those obtained from the stationary signal $u(i)$ and examine
how the scaling properties of a detrended fluctuation function $F(n)$ change
when introducing different types of nonstationarities.

\section{Signals with segments removed}\label{seccut}
 
\begin{figure}
\centerline{
\epsfysize=0.9\columnwidth{\rotate[r]{\epsfbox{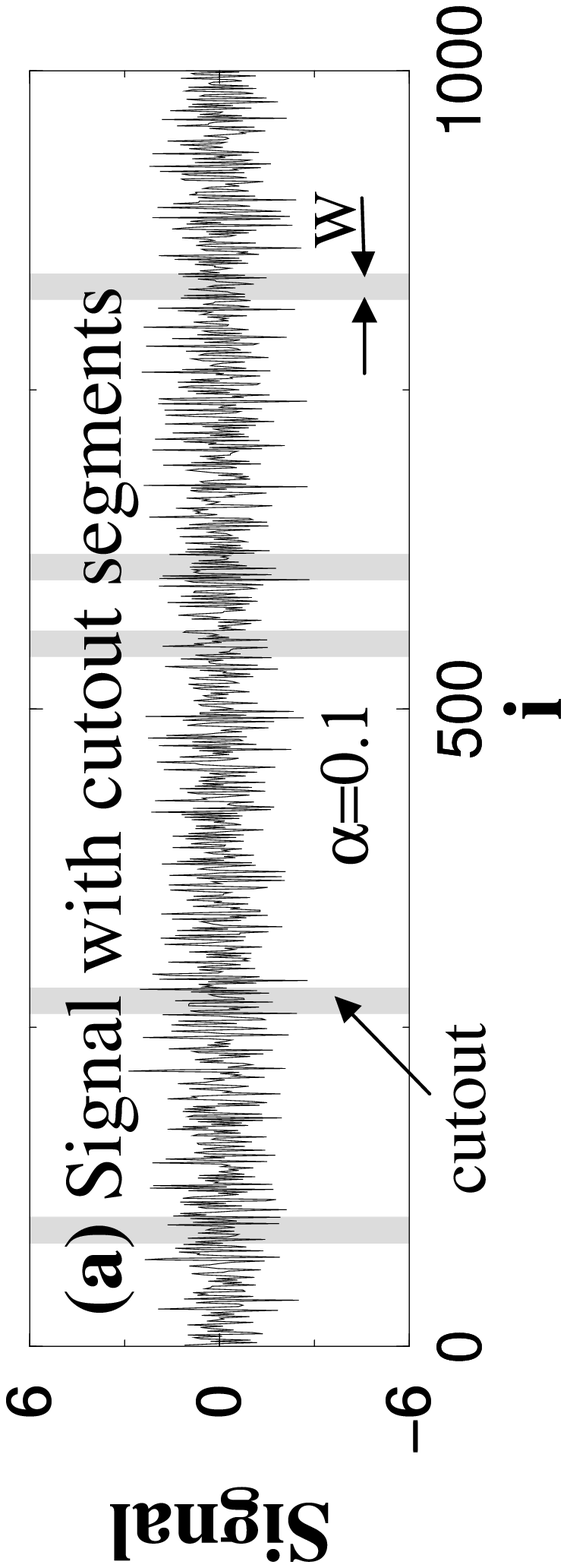}}}}
\vspace*{0.25cm}
\centerline{
\epsfysize=0.9\columnwidth{\rotate[r]{\epsfbox{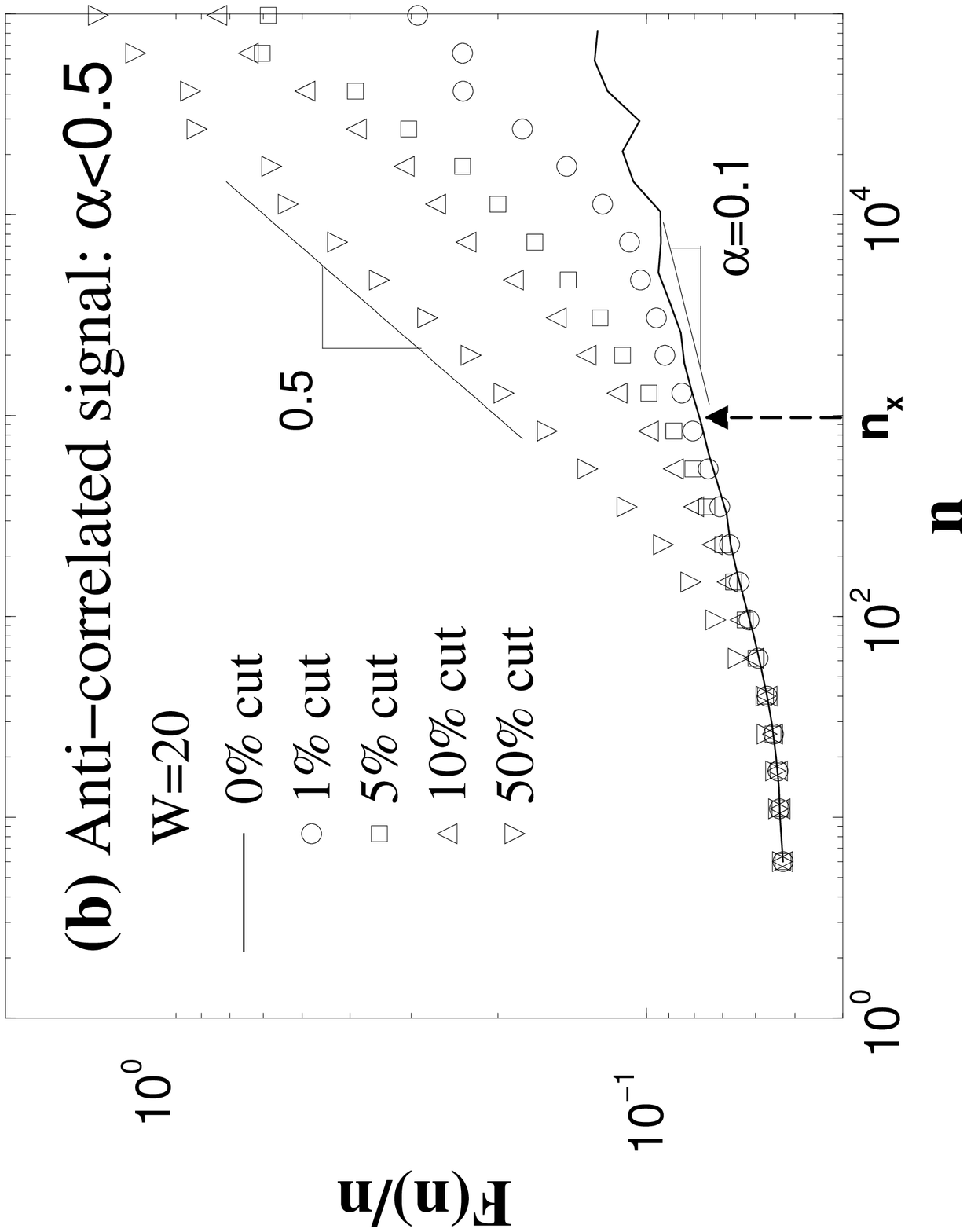}}}}

\centerline{
\epsfysize=0.9\columnwidth{\rotate[r]{\epsfbox{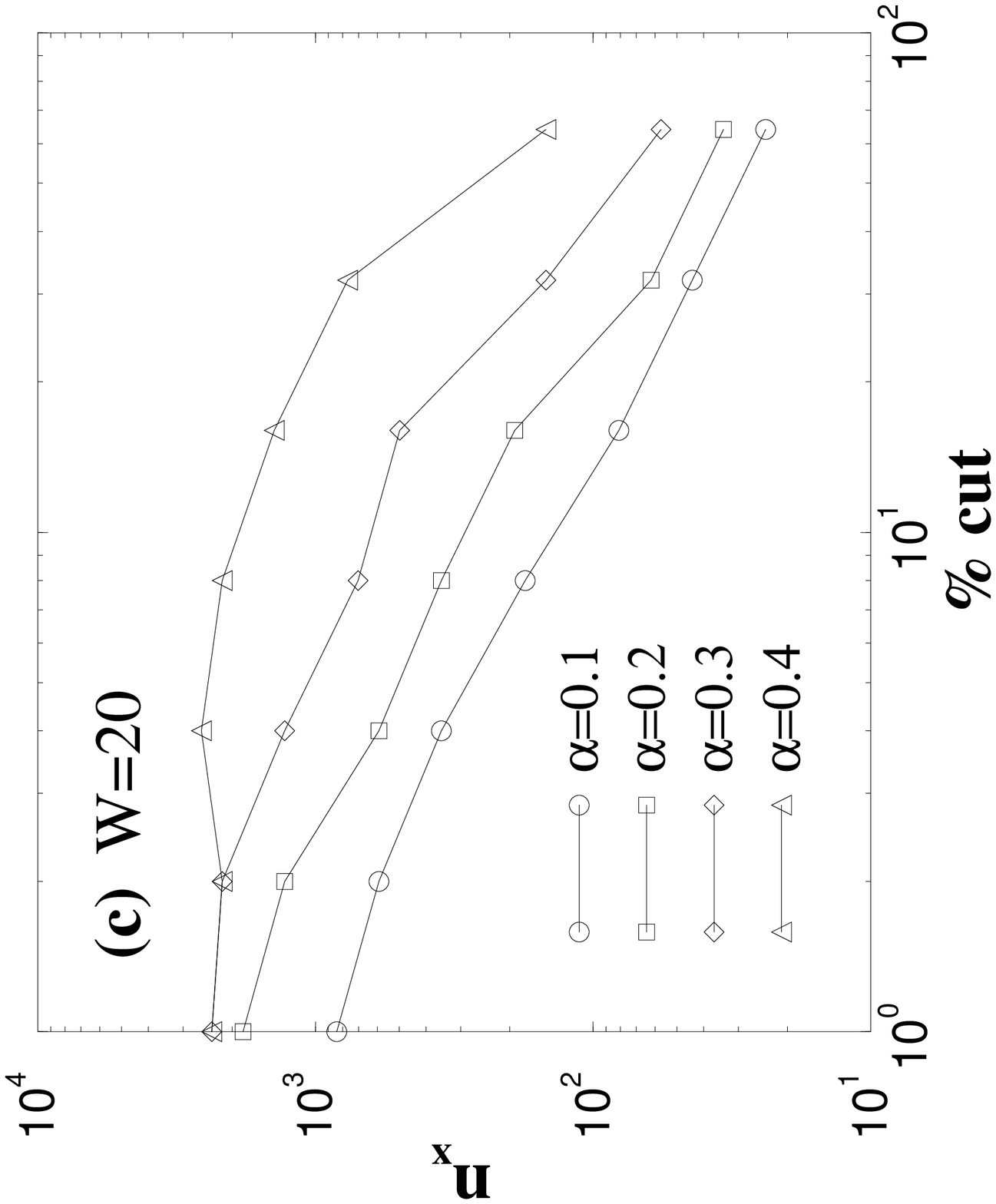}}}}
\centerline{
\epsfysize=0.9\columnwidth{\rotate[r]{\epsfbox{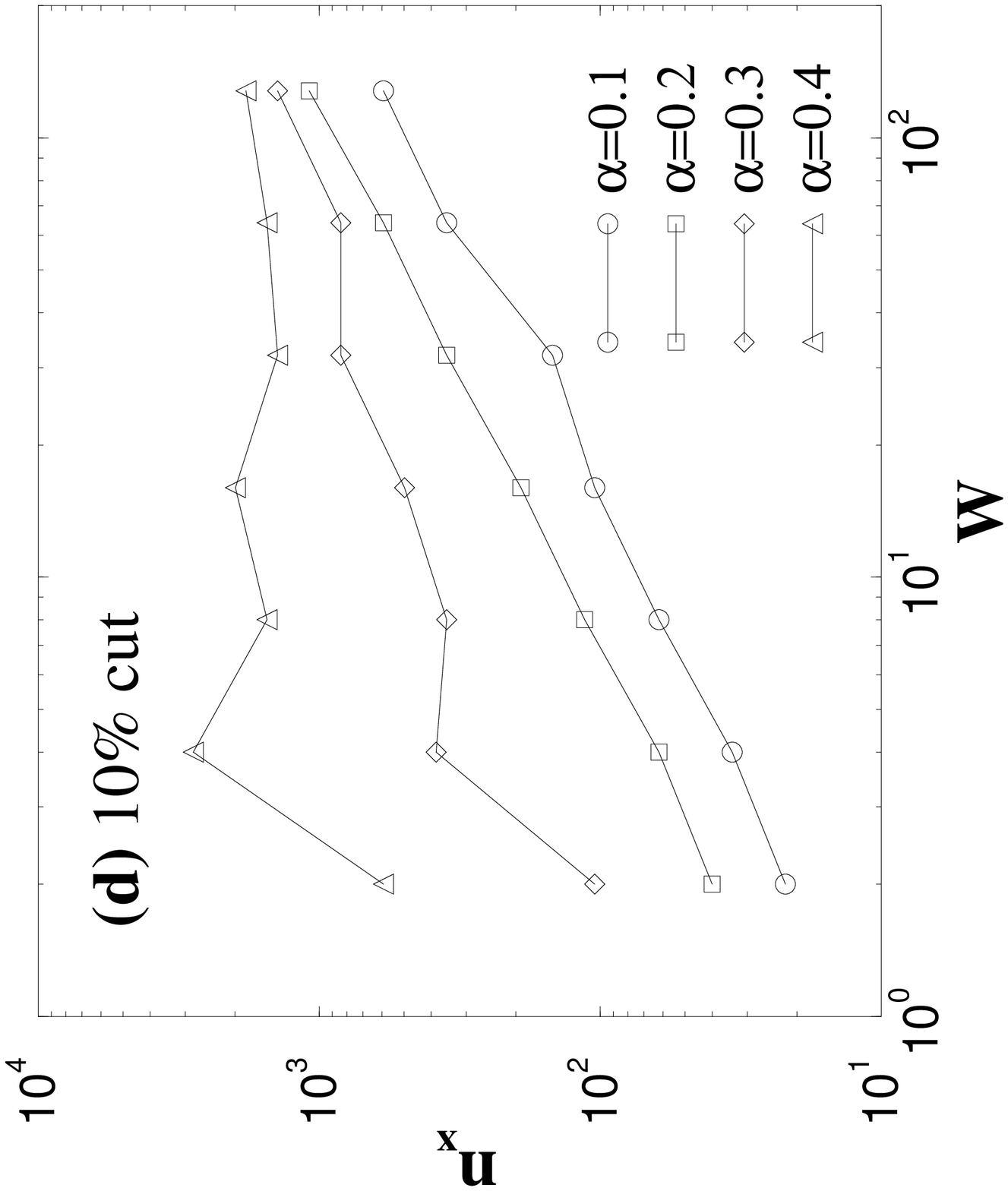}}}}
\centerline{
\epsfysize=0.9\columnwidth{\rotate[r]{\epsfbox{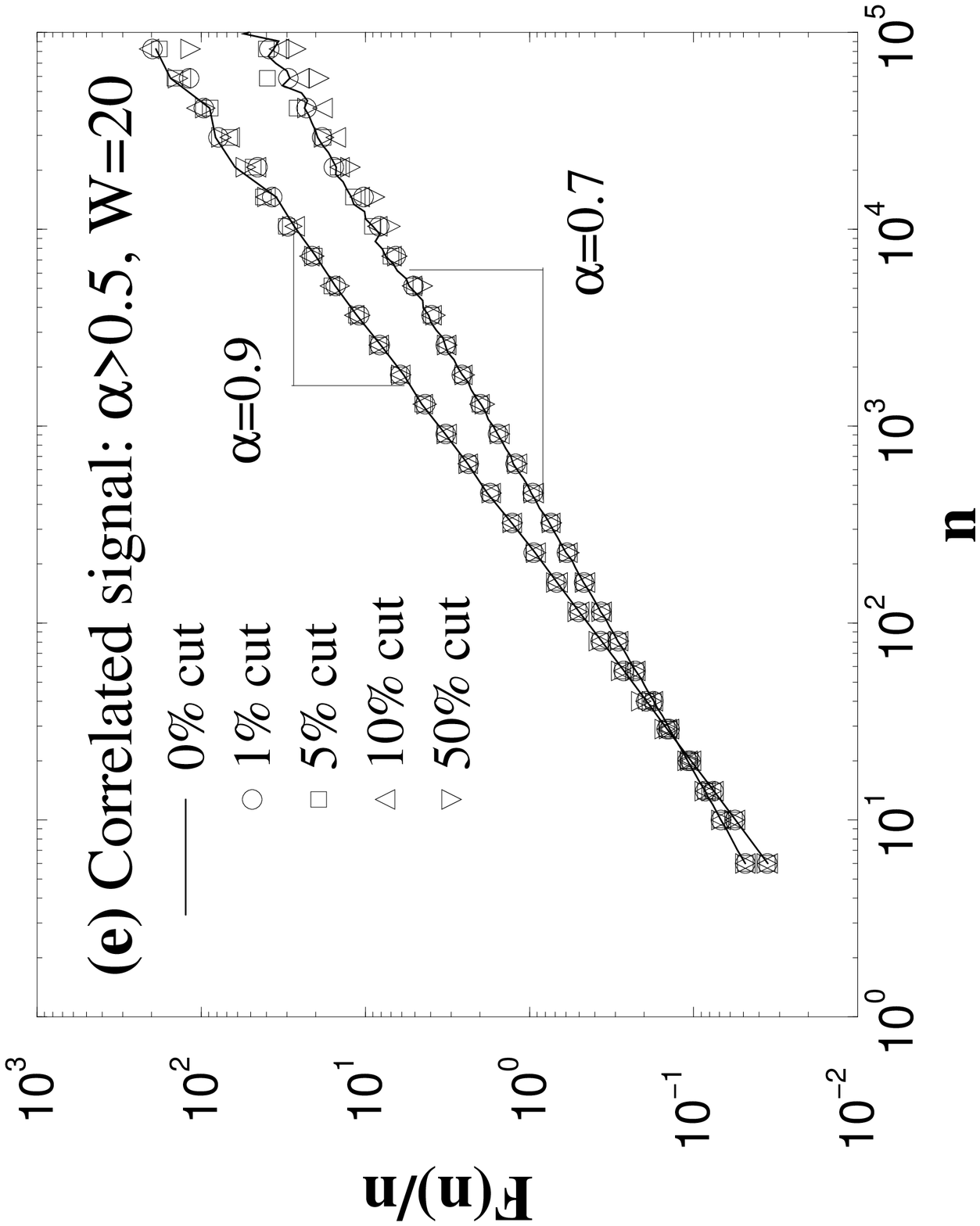}}}}

\caption{Effects of the ``cutting'' procedure on the scaling behavior of 
  stationary correlated signals. $N_{max}=2^{20}$ is the
  number of points in the signals (standard deviation $\sigma=1$) and $W$ is the
   size of the cutout segments. (a) A stationary signal with 10\% of the
  points removed. The removed parts are presented by shaded segments of size
  $W=20$ and the remaining parts are stitched together. (b) Scaling behavior
  of nonstationary signals obtained from an anti-correlated stationary signal
  (scaling exponent~$\alpha<0.5$) after the cutting procedure. A crossover
  from anti-correlated to uncorrelated ($\alpha=0.5$) behavior appears at
  scale $n_{\times}$. The crossover scale $n_{\times}$ decreases with   
  increasing the fraction of points removed from the signal. We determine
  $n_{\times}$ based on the difference $\Delta$ between the logarithm of
  $F(n)/n$ for the original stationary anti-correlated signal ($\alpha=0.1$)
  and the nonstationary signal with cutout segments: $n_{\times}$ is
  the scale at which $\Delta\geq0.04$. Dependence of the crossover scale
  $n_{\times}$ on the fraction (c) and on the size $W$ (d) of the cutout
  segments for anti-correlated signals with different scaling exponent
  $\alpha$. (e) Cutting procedure applied to correlated signals
  ($\alpha>0.5$). In contrast to (b), no discernible effect on the scaling
  behavior is observed for different values of the scaling exponent $\alpha$,
  even when up to 50\% of the points in the signals are removed.
}
\label{cut}
\end{figure}

In this section, we study the effect of nonstationarity caused by removing
 segments of a given length from a signal and stitching together the remaining 
parts --- a ``cutting'' procedure often used in pre-processing data prior to 
analysis. To address this question, we first generate a stationary correlated
 signal $u(i)$ (see Sec.~\ref{secpuren}) of length $N_{max}$ 
and a scaling exponent $\alpha$, using the
modified Fourier filtering method\cite{MFFM}. Next, we divide this signal into
$N_{max}/W$ non-overlapping segments of size $W$ and randomly remove
some of these segments. Finally, we stitch together the remaining segments in  
the signal $u(i)$ [Fig.~\ref{cut}(a)], thus obtaining a surrogate 
nonstationary signal which is characterized by three parameters: the scaling 
exponent $\alpha$, the segment size $W$ and the fraction of the signal 
$u(i)$, which is removed.  

We find that the scaling behavior of such a nonstationary 
 signal strongly depends on the scaling exponent $\alpha$ of the original
 stationary correlated signal $u(i)$. As illustrated in  Fig.~\ref{cut}(b), 
for a stationary {\it anti-correlated\/} signal with $\alpha=0.1$, the
 cutting procedure
 causes a crossover in the scaling behavior of the resultant nonstationary
 signal. This crossover appears even when only $1\%$ of the
 segments are cut out. At the scales larger than the crossover scale
 $n_{\times}$ the r.m.s.~fluctuation function  behaves 
 as $F(n) \sim n^{0.5}$, which means an uncorrelated randomness, i.e.,  the
 anti-correlation has been completely destroyed in this regime. For all
 anti-correlated signals with exponent $\alpha<0.5$, we observe a  
 similar crossover behavior. This result is surprising, since researchers often
 take for granted that a cutting procedure before analysis does not
 change the scaling properties of the original signal. Our simulation shows that
 this assumption is not true, at least for anti-correlated signals.   
 
Next, we investigate how the two parameters --- the segment size $W$ and the
fraction of points cut out from the signal --- control the effect of the
cutting procedure on the scaling behavior of anti-correlated signals. For
the fixed size of the segments ($W=20$), we find that the crossover scale
$n_{\times}$ {\it decreases} with {\it increasing} the fraction of the
cutout segments [Fig.~\ref{cut}(c)]. Furthermore, for anti-correlated
signals with small values of the scaling exponent $\alpha$, e.g.,
$\alpha=0.1$ and $\alpha=0.2$, we find that $n_{\times}$ and the fraction of
the cutout segments display an approximate power-law relationship. For a
fixed fraction of the removed segments, we find that the crossover scale
$n_{\times}$ {\it increases} with {\it increasing} the segment size $W$
[Fig.~\ref{cut}(d)]. To minimize the effect of the cutting procedure on the
correlation properties, it is advantageous to cut smaller number of segments
of larger size $W$.~Moreover, if the segments which need to be removed are
too close (e.g., at a distance shorter than the size of the segments), it
may be advantageous to cut out both the segments and a                       part of the signal
between them. This will effectively increase the size of the segment $W$
without substantially changing the fraction of the signal which is cut out,
leading to an increase in the crossover scale $n_{\times}$. Such strategy
would minimize the effect of this type of nonstationarity on the scaling
properties of data. For small values of the scaling exponent $\alpha$
($\alpha<0.25$), we find that $n_{\times}$ and $W$ follow power-law
relationships [Fig.~\ref{cut}(d)]. The reason we do not observe a power-law
relationship between $n_{\times}$ and $W$ and between $n_{\times}$ and the
fraction of cutout segments for the values of the scaling exponent $\alpha$
close to $0.5$ may be due to the fact that the crossover regime becomes
broader when it separates scaling regions with similar exponents, thus
leading to uncertainty in defining $n_{\times}$. For a fixed $W$ and a fixed
fraction of the removed segments [see Figs.~\ref{cut}(c) and (d)], we
observe that $n_{\times}$ increases with the increasing value of the scaling
exponent $\alpha$, i.e., the effect of the cutting procedure on the scaling
behavior decreases when the anti-correlations in the signal are weaker
($\alpha$ closer to $0.5$).

Finally, we consider the case of correlated signals $u(i)$ with 
$1.5>\alpha>0.5$. Surprisingly, we find that the scaling of correlated signals
is not affected by the cutting procedure. This observation remains true
independently of the segment size $W$ --- from very small $W=5$ up to very
large $W=5000$ segments --- even when up to $50\%$ of the segments are
removed from a signal with $N_{max}\sim 10^{6}$ points [Fig.~\ref{cut}(e)].

\section{Signals with random spikes}\label{secrspikes}

\begin{figure}  
\centerline{
\epsfysize=0.9\columnwidth{\rotate[r]{\epsfbox{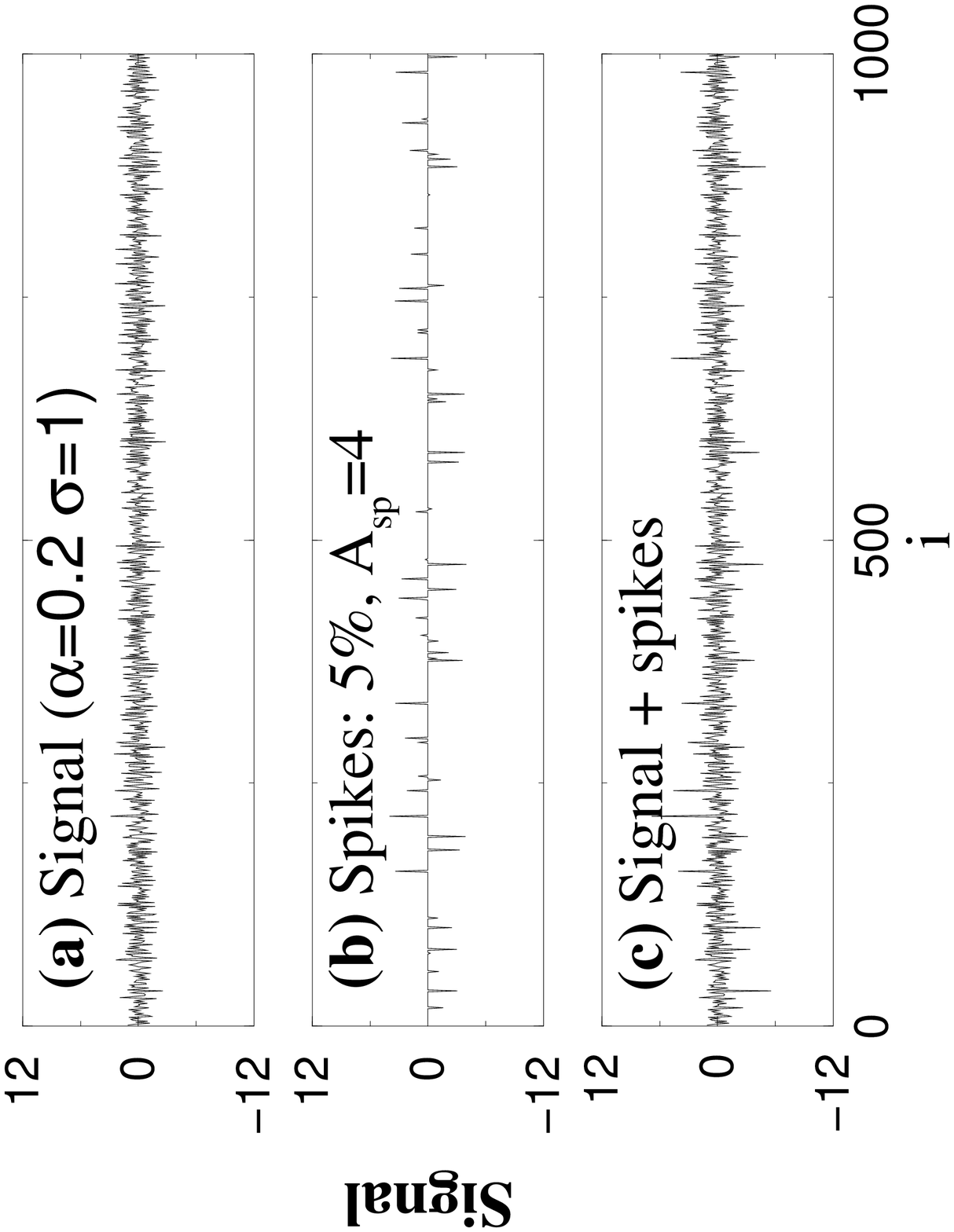}}}}
\centerline{
\epsfysize=0.9\columnwidth{\rotate[r]{\epsfbox{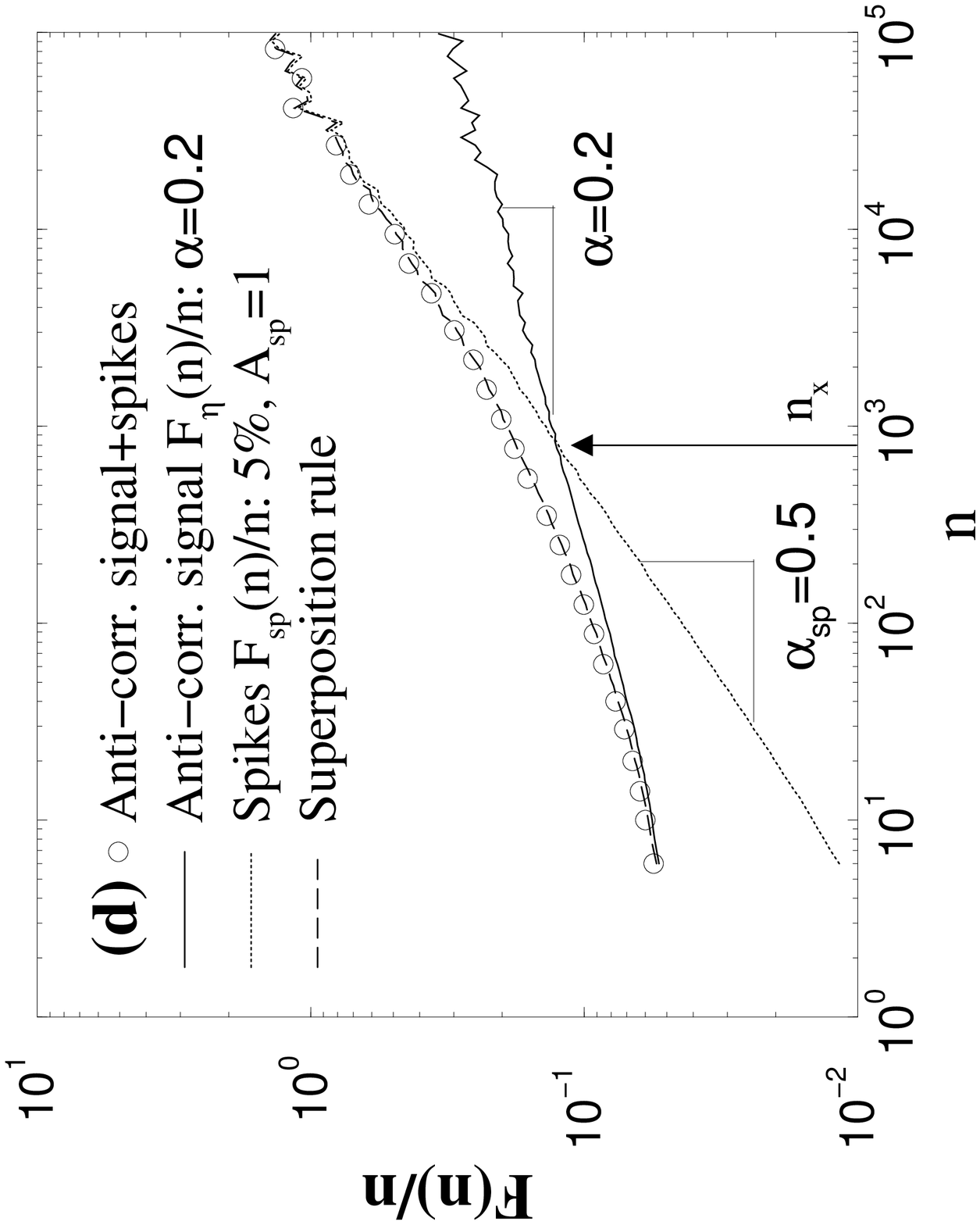}}}}
\centerline{
\epsfysize=0.9\columnwidth{\rotate[r]{\epsfbox{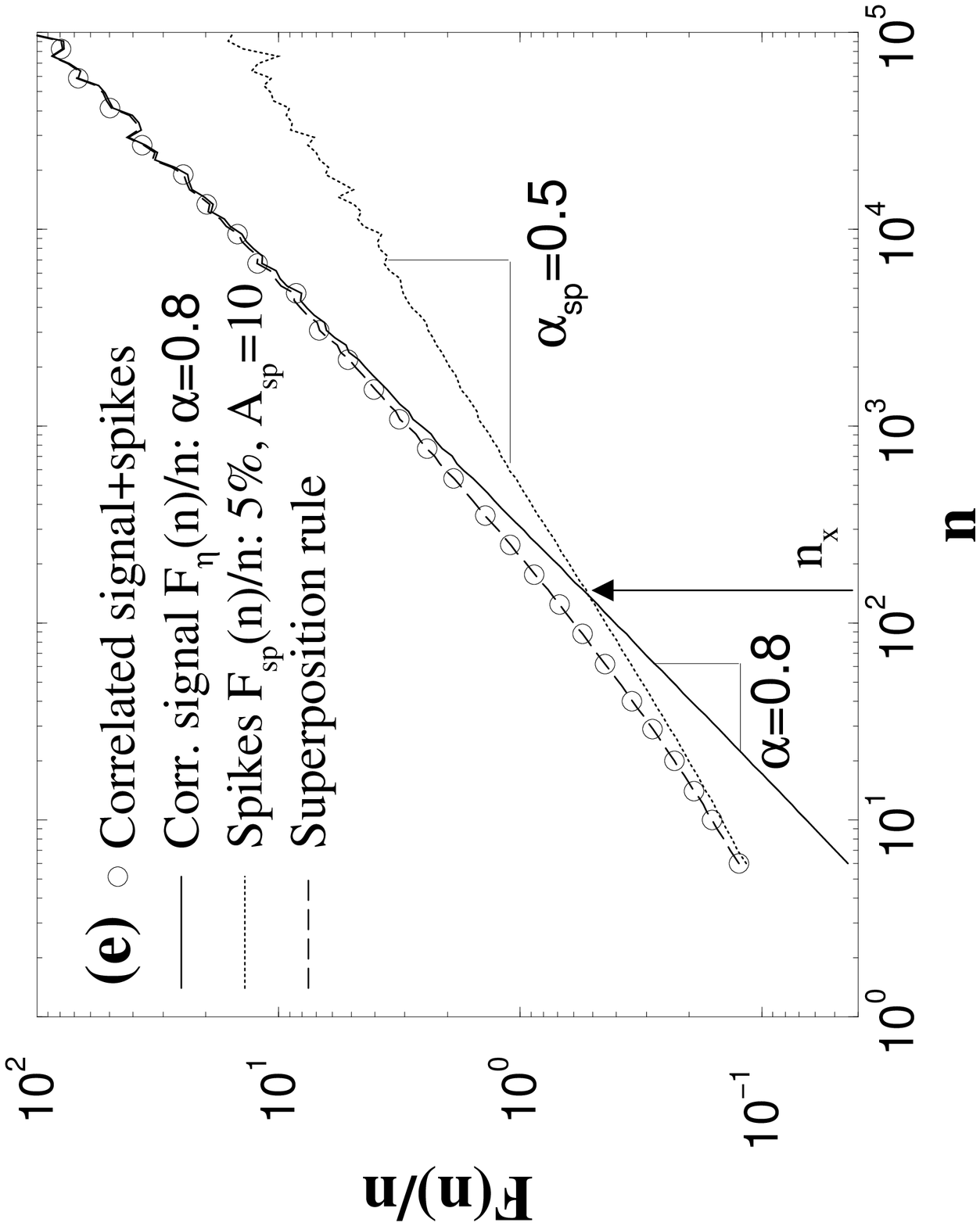}}}} 
\caption{Effects of random spikes on the scaling behavior of stationary 
correlated signals. (a) An example of an anti-correlated signal $u(i)$  with 
scaling exponent $\alpha=0.2$,  $N_{max}=2^{20}$ 
and standard deviation $\sigma=1$. (b) A series of uncorrelated spikes 
($\alpha_{sp}=0.5$) at 5$\%$ randomly chosen positions (concentration
$p=0.05$) and with uniformly distributed amplitudes $A_{sp}$ in the
interval $[-4, 4]$. (c) The superposition of the signals in 
(a) and (b). (d) Scaling behavior of an anti-correlated signal $u(i)$
($\alpha=0.2$) with spikes~($A_{sp}=1$, $p=0.05$, $\alpha_{sp}=0.5$). For
$n<n_{\times}$, $F(n)/n \approx F_{\eta}(n)/n \sim n^{\alpha}$, where
$F_{\eta}(n)/n$ is the scaling function of the signal $u(i)$. For $n>n_{\times}$, 
$F(n)/n \approx F_{sp}(n)/n \sim n^{\alpha_{sp}}$. (e) Scaling behavior of a 
correlated signal $u(i)$ ($\alpha=0.8$) with 
spikes ($A_{sp}=10$, $p=0.05$, $\alpha_{sp}=0.5$). For $n<n_{\times}$, 
$F(n)/n \approx F_{sp}(n)/n \sim n^{\alpha_{sp}}$. For $n>n_{\times}$, 
$F(n)/n \approx F_{\eta}(n)/n \sim n^{\alpha}$. Note that when 
$\alpha=\alpha_{sp}=0.5$, there is no crossover. }
\label{spikes} 
\end{figure}

In this section, we consider nonstationarity related to
the presence of random spikes in data and we study the effect of this type
of nonstationarity on the scaling properties of correlated signals. First, we
generate surrogate nonstationary signals by adding random spikes
 to a stationary correlated signal $u(i)$ [see Sec.~\ref{secpuren} and
 Fig.~\ref{spikes}(a-c)].

We find that the correlation properties of the nonstationary signal with
spikes depend on the scaling exponent $\alpha$ of the stationary signal and
the scaling exponent $\alpha_{sp}$ of the spikes. 
When uncorrelated spikes ($\alpha_{sp}=0.5$) are added to a correlated
or anti-correlated stationary signal [Fig~\ref{spikes}(d) and (e)], we
observe a change in the scaling behavior with a crossover at a characteristic
 scale $n_{\times}$. For anti-correlated signals ($\alpha<0.5$) with random
 spikes, we find that at scales smaller than $n_{\times}$, the scaling
 behavior is close to the one observed for the stationary anti-correlated
 signal without spikes, while for scales larger than $n_{\times}$, there is a
 crossover to random behavior. In the case of correlated signals
 ($\alpha>0.5$) with random spikes, we find a different crossover from
 uncorrelated behavior at small scales, to correlated behavior at large
 scales with an exponent close to the exponent of the original stationary
 correlated signal. Moreover, we find that spikes with a very small amplitude
 can cause strong crossovers in the case of anti-correlated signals, while
 for correlated signals, identical concentrations of spikes with a much larger
 amplitude do not affect the scaling.
Based on these findings, we conclude that uncorrelated spikes with a 
sufficiently large amplitude can affect the DFA results at large scales for
signals with $\alpha<0.5$ and at small scales for signals with $\alpha>0.5$.

To better understand the origin of this crossover behavior, we first study the
scaling of the spikes only [see Fig.~\ref{spikes}(b)]. By varying
the concentration $p$ ($0\leq p\leq 1$) and the amplitude $A_{sp}$ of the
spikes in the signal, we find that for the general case when the spikes may
be correlated, the r.m.s. fluctuation function behaves as 
\begin{equation}
F_{sp}(n)/n=k_0\sqrt{p}A_{sp}n^{\alpha_{sp}},
\label{si_bl0}
\end{equation}  
where $k_0$ is a constant and $\alpha_{sp}$ is the scaling exponent of the 
 spikes. 

Next, we investigate the analytical relation between the DFA results 
obtained from the original correlated signal, the spikes and the
superposition of signal and spikes. Since the original signal and the spikes
are not correlated, we can use a {\it superposition rule\/} (see \cite{kun}
and Appendix~\ref{secadd}) to derive the
r.m.s. fluctuation function $F(n)/n$ for the correlated signal with spikes: 
\begin{equation}
[F(n)/n]^2=[F_{\eta}(n)/n]^2+[F_{sp}(n)/n]^2,
\label{mixadd1}
\end{equation}
where $F_{\eta}(n)/n$ and $F_{sp}(n)/n$ are the r.m.s.~fluctuation function for
the signal and the 
spikes, respectively. To confirm this theoretical result, we calculate
$\sqrt{[F_{\eta}(n)/n]^2+[F_{sp}(n)/n]^2}$ [see Figs.~\ref{spikes}(d),
(e)] and find this Eq.~(\ref{mixadd1}) is remarkably consistent
with our experimental observations. 

Using the superposition rule, we can also theoretically predict the crossover
scale $n_{\times}$ as the intercept between $F_{\eta}(n)/n$ and  $F_{sp}(n)/n$,
i.e., where $F_{\eta}(n_{\times})=F_{sp}(n_{\times})$.  We find that 
\begin{equation}
n_{\times} = \left(\sqrt{p}A_{sp}\frac{k_0}{b_0}\right)^{1/(\alpha-\alpha_{sp})},
\label{si_bl}
\end{equation}
since the r.m.s. fluctuation function for the signal and the spikes are
$F_{\eta}(n)/n=b_0 n^{\alpha}$ \cite{kun} and
$F_{sp}(n)/n=k_0\sqrt{p}A_{sp}n^{\alpha_{sp}}$ [Eq.~(\ref{si_bl0})], 
respectively. This result predicts the position of the crossover depending on
the parameters defining the signal and the spikes. 

Our result derived from the superposition rule can be useful to distinguish
two cases:  
({\it i}) the correlated stationary signal and the spikes are independent (e.g., the
case when a correlated signal results from the intrinsic dynamics of the
system while the spikes are due to external perturbations); and ({\it ii}) the
correlated stationary signal and the spikes are dependent (e.g., both the
signal and the spikes arise from the intrinsic dynamics of the system). In the
latter case, the identity in the superposition rule is not correct (see
Appendix~\ref{secadd}).  

\section{Signals with different local behavior}\label{secmix}

Next, we study the effect of nonstationarities on complex patchy signals where 
different segments show different local behavior. This type of
nonstationarity is very common in real world 
data\cite{cknature1992,SMDFA1,plamennature1996,bundesleep2000,SLEEP1}. Our discussion of signals composed of
only two types of segments is limited to two simple cases: (A) different
standard deviations and (B) different correlations. 
   
\subsection{Signals with different local standard deviations}

\begin{figure}
\centerline{
\epsfysize=0.9\columnwidth{\rotate[r]{\epsfbox{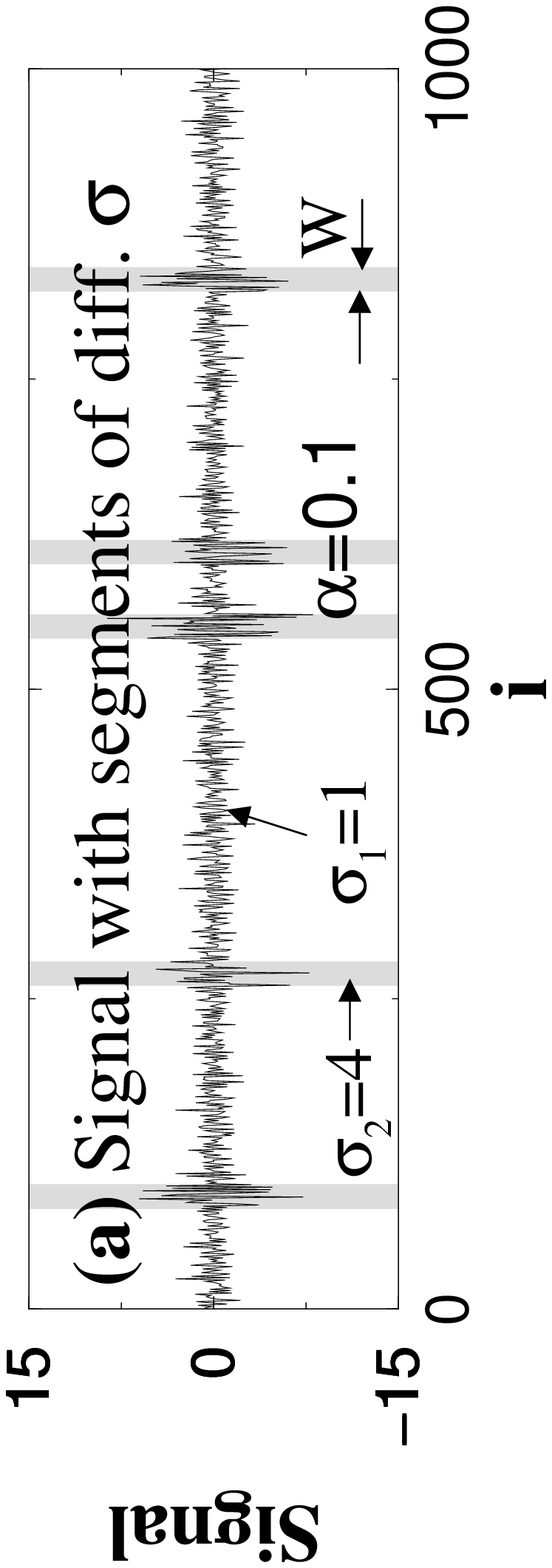}}}}
\vspace*{0.25cm}
\centerline{
\epsfysize=0.9\columnwidth{\rotate[r]{\epsfbox{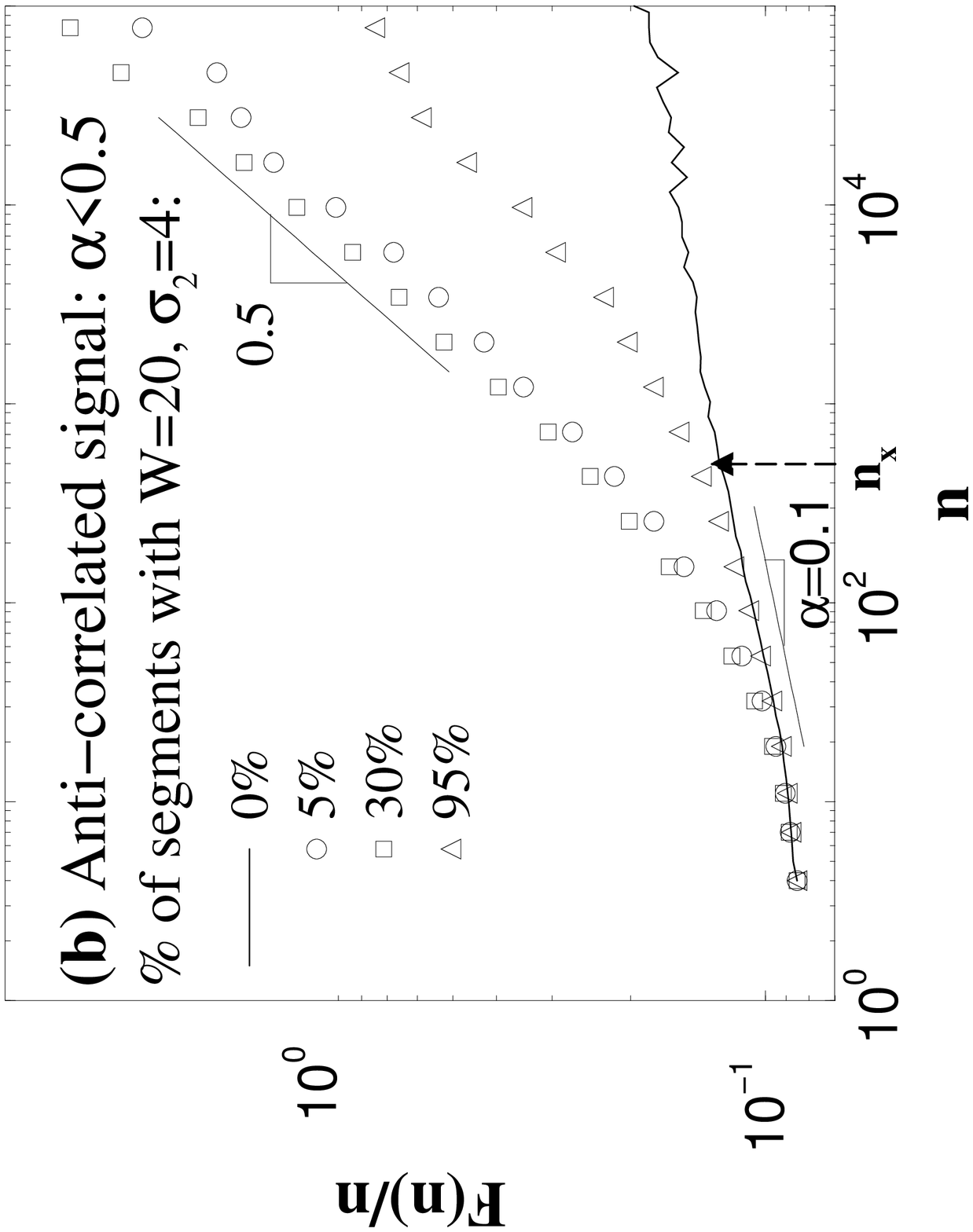}}}}
\centerline{
\epsfysize=0.9\columnwidth{\rotate[r]{\epsfbox{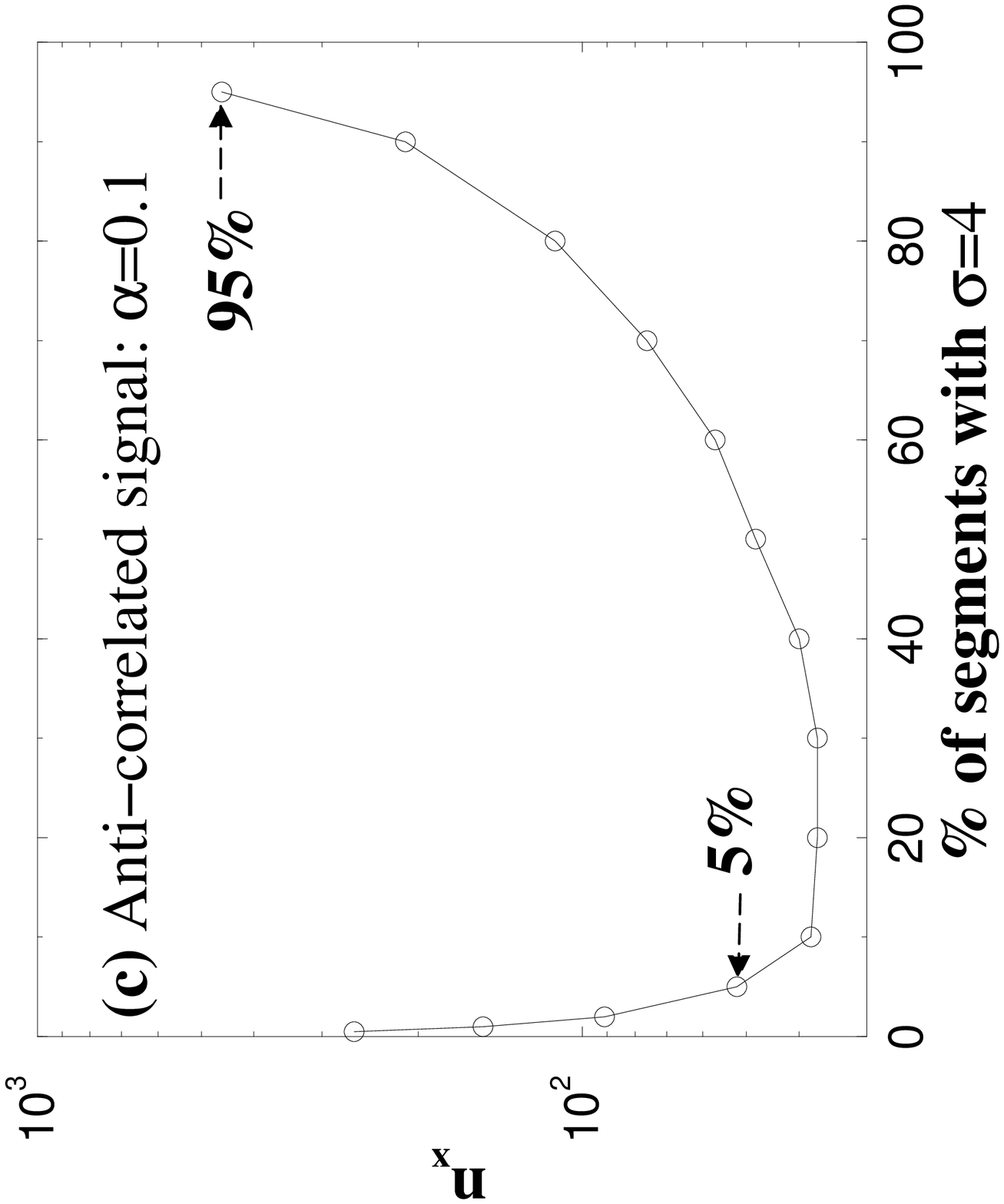}}}}
\centerline{
\epsfysize=0.9\columnwidth{\rotate[r]{\epsfbox{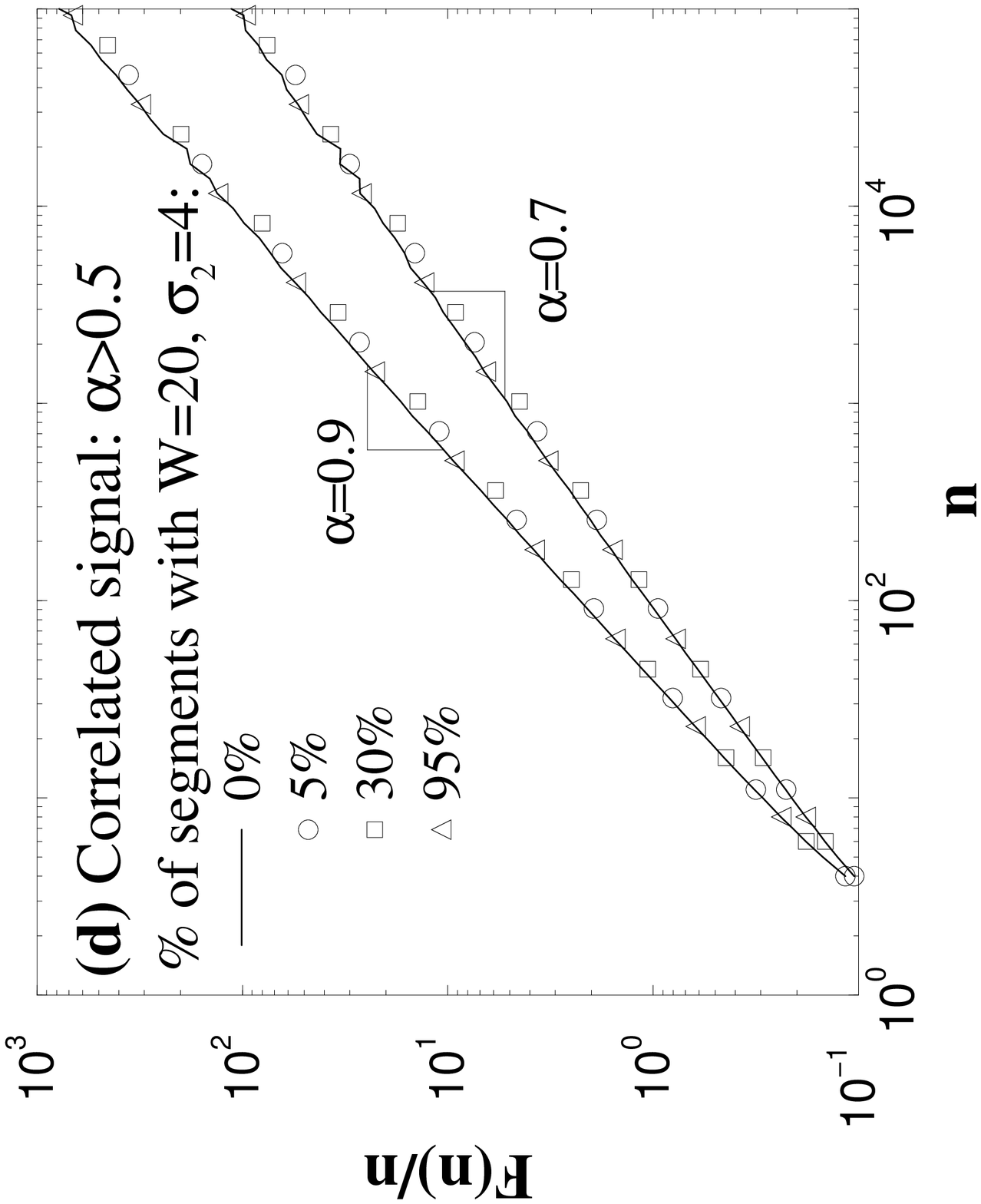}}}}
\caption{Scaling behavior of nonstationary correlated signals with different
  local standard deviation. (a)
  Anti-correlated signal ($\alpha=0.1$) with standard deviation $\sigma_1=1$
  and amplified segments with standard deviation $\sigma_2=4$. The size of
  each segment is $W=20$ and the fraction of the amplified segments is
  $p=0.1$ from the total length of the signal ($N_{max}=2^{20}$). (b) Scaling
  behavior of the signal in (a) for a different fraction $p$ of the amplified
  segments (after normalization of the globe standard deviation to unity). A
  crossover from anti-correlated behavior ($\alpha=0.1$) at small scales to
  random behavior ($\alpha=0.5$) at large scales is observed. (c) Dependence
  of the crossover scale $n_{\times}$ on the fraction $p$ of amplified
  segments for the signal in (a). $n_{\times}$ is determined from the 
  difference $\Delta$ of $\log_{10}[F(n)/n]$ between the nonstationary signal
  with amplified segments and the original stationary signal. Here we choose
  $\Delta=0.04$. (d) Scaling behavior of nonstationary signals obtained from
  correlated stationary signals ($1>\alpha>0.5$) with standard deviation
  $\sigma_1=1$, for a different fraction of the amplified segments with
  $\sigma_2=4$. No difference in the scaling is observed, compared to the
  original stationary signal.} 
\label{mixamp}
\end{figure}

Here we consider nonstationary signals comprised of segments with the same 
local scaling exponent, but different local standard deviations. We first 
generate a stationary correlated signal $u(i)$ (see Sec.~\ref{secpuren}) with
fixed standard deviation  
$\sigma_1=1$. Next, we divide the signal $u(i)$ into non-overlapping 
segments of size $W$. We then randomly choose a fraction $p$ of the 
segments and amplify the standard deviation of the signal in these 
segments, $\sigma_2=4$ [Fig.\ref{mixamp}(a)]. Finally, we 
 normalize the entire signal to global standard deviation $\sigma=1$
 by dividing the value of each point of the signal by  
$\sqrt{(1-p)\sigma_1^2+p\sigma_2^2}$. 

For nonstationary {\it anti-correlated \/} signals ($\alpha<0.5$) with 
segments characterized by two different values of the standard deviation, we 
observe a crossover at scale $n_{\times}$ [Fig.\ref{mixamp}(b)]. For small 
scales $n<n_{\times}$, the behavior is anti-correlated with an exponent equal
 to the scaling exponent $\alpha$ of the original stationary anti-correlated
 signal $u(i)$. For large scales $n>n_{\times}$, we find a transition to random 
 behavior with exponent $0.5$, indicating that the anti-correlations have 
been destroyed. The dependence of crossover scale 
$n_{\times}$ on the fraction $p$ of segments with larger standard deviation 
is shown in Fig.~\ref{mixamp}(c). The dependence is not monotonic because 
for $p=0$ and $p=1$, the local standard deviation is constant throughout the 
signal, i.e., the signal becomes stationary and thus there is no crossover. 
Note the asymmetry in the value of $n_{\times}$ --- a much smaller value of 
$n_{\times}$ for $p=0.05$ compared to $p=0.95$ [see Fig.~\ref{mixamp}(b-c)].
 This result indicates that very few segments with a large standard deviation 
(compared to the rest of the signal) can have a strong effect on the
anti-correlations in the signal. Surprisingly, the same fraction of segments
with a small standard deviation (compared to the rest of the signal) does not
affect the anti-correlations up to relatively large scales. 

For nonstationary {\it correlated\/} signals ($\alpha>0.5$) with 
segments characterized by two different values of the standard deviation, 
we surprisingly find no difference in the scaling of $F(n)/n$, compared to 
 the stationary correlated signals with constant standard deviation 
[Fig.~\ref{mixamp}(d)]. Moreover, this observation remains valid for different
 sizes of the segments $W$ and for different values of the fraction $p$ of 
segments with larger standard deviation. We note that in the limiting case of 
very large values of $\sigma_{2}/\sigma_{1}$, when the values of the signal 
in the segments with standard deviation $\sigma_{1}$ could be considered 
close to ``zero'', the results in Fig.~\ref{mixamp}(d) do not hold and we 
observe a scaling behavior similar to that of the signal in 
Fig.~\ref{mix}(c) (see following section).

\subsection{Signals with different local correlations}

\begin{figure}
\centerline{
\epsfysize=0.9\columnwidth{\rotate[r]{\epsfbox{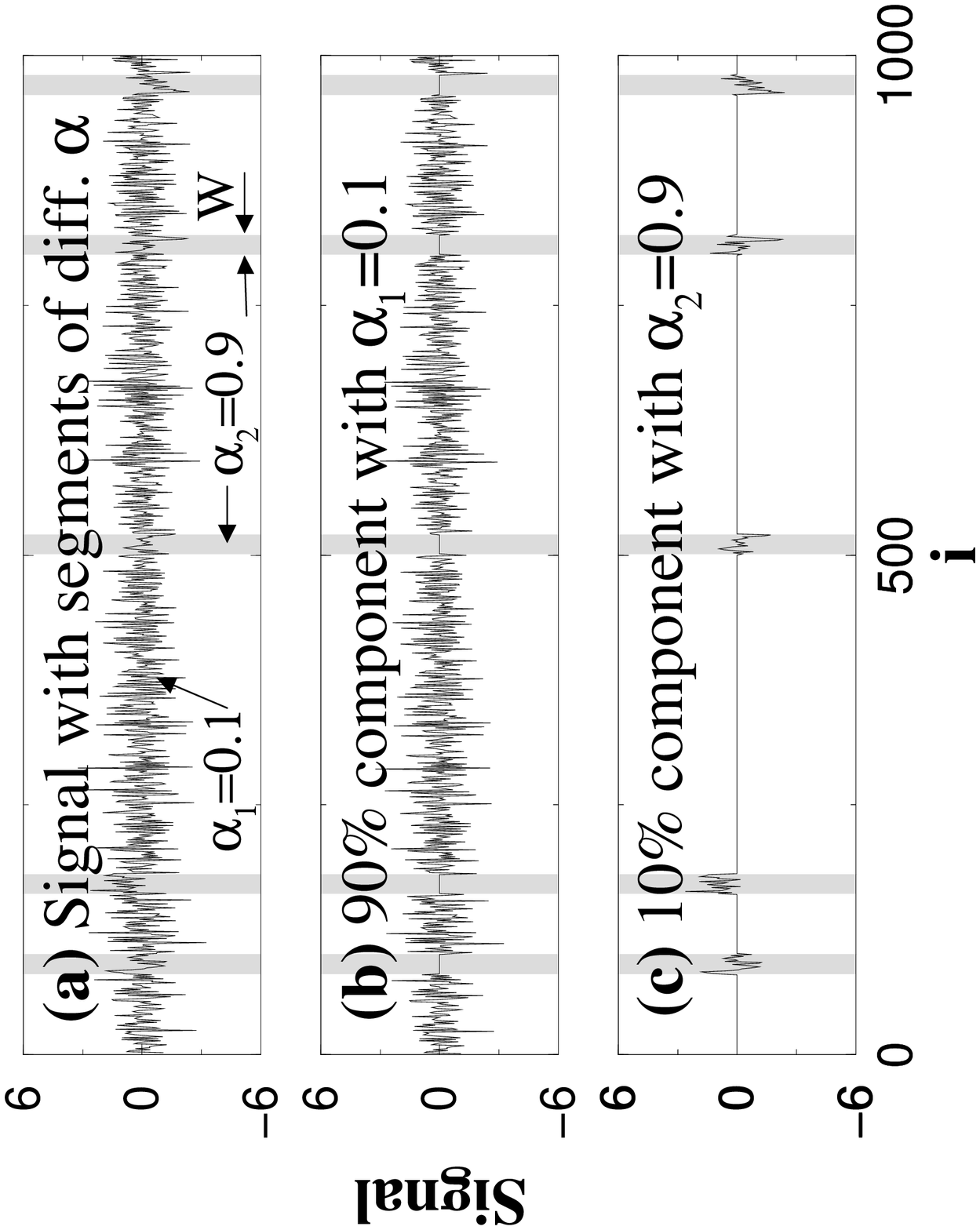}}}}
\centerline{
\epsfysize=0.9\columnwidth{\rotate[r]{\epsfbox{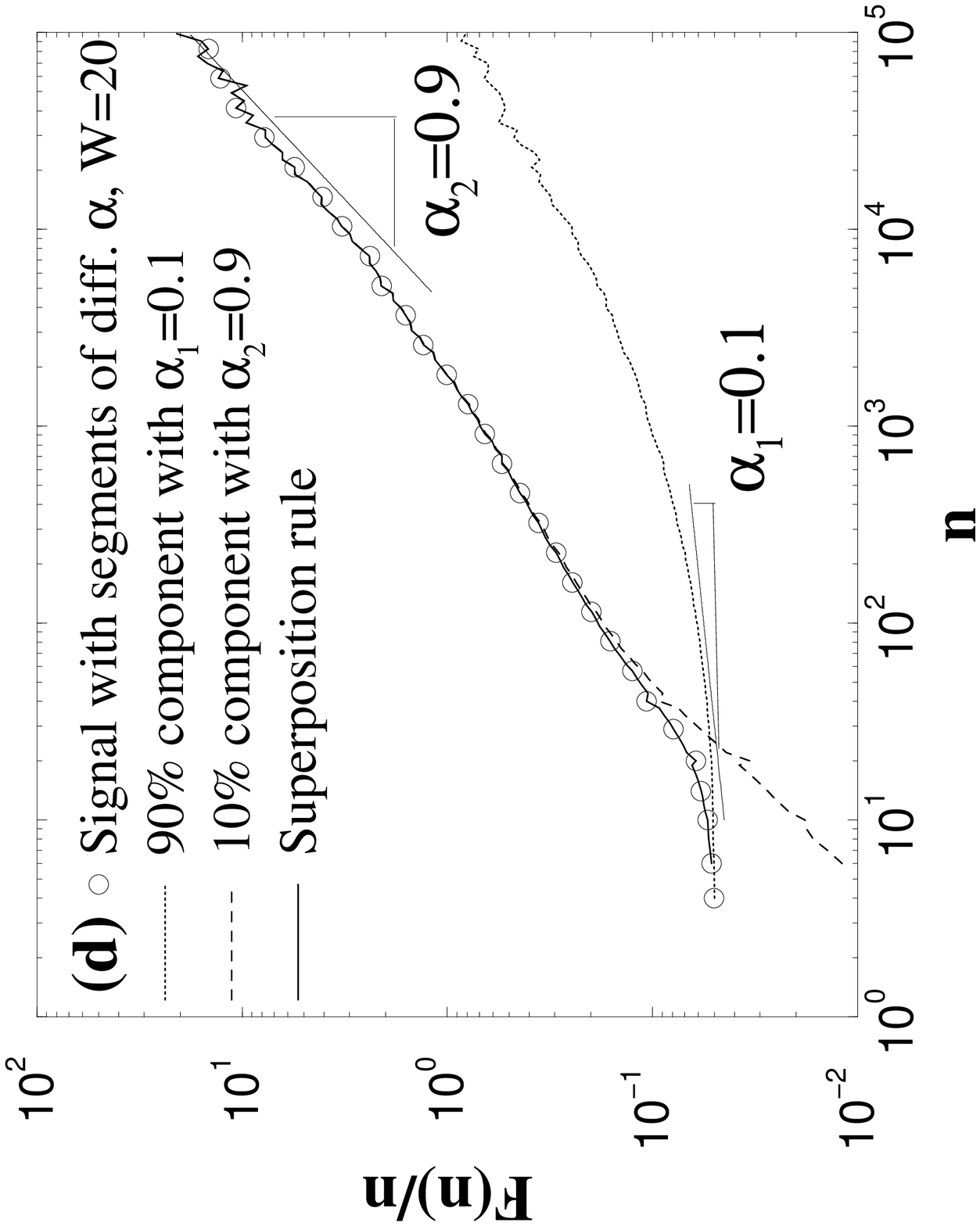}}}}
\vspace*{0.25cm}
\caption{Scaling behavior of a nonstationary signal with two different
  scaling exponents. (a) Nonstationary signal (length $N_{max}=2^{20}$,
  standard deviation $\sigma=1$) which is a mixture of correlated segments
  with exponent $\alpha_{1}=0.1$ (90\% of the signal) and segments with
  exponent $\alpha_{2}=0.9$ (10\% of the signal). The segment size is $W=20$;
  (b) the 90\% component containing all segments with $\alpha_{1}=0.1$ and
  the remaining segments (with $\alpha_{2}=0.9$) are replaced by zero; (c)
  the 10\% component containing all segments with $\alpha_{2}=0.9$ and 
  the remaining segments (with $\alpha_{1}=0.1$) are replaced by zero; (d) DFA
  results for the mixed signal in (a), for the individual components in
  (b) and (c), and our prediction obtained from the superposition rule.} 
\label{mix}
\end{figure}

Next we consider nonstationary signals which consist of segments with
identical standard deviation ($\sigma=1$) but different correlations. We
obtain such 
signals using the following procedure: (1) we generate two stationary signals
$u_{1}(i)$ and $u_{2}(i)$ (see Sec.~\ref{secpuren}) of identical length
$N_{max}$ and with different correlations, characterized by scaling exponents
$\alpha_{1}$ and $\alpha_{2}$; (2) we divide the signals $u_{1}(i)$ and
$u_{2}(i)$ into non-overlapping segments of size $W$; (3) we randomly replace
a fraction $p$ of the segments in signal $u_{1}(i)$ with the corresponding
segments of $u_{2}(i)$. In Fig.~\ref{mix}(a), we show an example of such a  
complex nonstationary signal with different local correlations. In this
Section, we study the behavior of the r.m.s. fluctuation function
$F(n)/n$. We also investigate $F(n)/n$ separately for each 
component of the nonstationary signal (which consists only of the segments
with identical local correlations) and  suggest an approach, based on the
DFA results, to recognize such complex structures in real data.

In  Fig.~\ref{mix}(d), we present the DFA result on such a nonstationary
signal, composed of segments with two different types of local correlations
characterized by exponents $\alpha_{1}=0.1$ and $\alpha_{2}=0.9$. We find
that at small scales, the slope of $F(n)/n$ is close to $\alpha_{1}$ and at
large scales, the slope approaches $\alpha_{2}$ with a bump in the
intermediate 
scale regime. This is not surprising, since $\alpha_{1}<\alpha_{2}$ and thus
$F(n)/n$ is bound to have a small slope ($\alpha_{1}$) at small scales and 
a large slope ($\alpha_{2}$) at large scales. However, it is surprising that
although 90\% of the signal consists of segments with scaling exponent
$\alpha_{1}$, $F(n)/n$ deviates at small scales ($n\approx 10$) from the
behavior expected for an anti-correlated signal $u(i)$ with exponent
$\alpha_{1}$ [see, e.g., 
the solid line in Fig.~\ref{cut}(b)]. This suggests that the behavior of
$F(n)/n$ for a nonstationary signal comprised of mixed segments with different
correlations is dominated by the segments exhibiting higher positive
correlations even in the case when their relative fraction in the signal is
 small. This observation is pertinent to real data such
as: ({\it i}) heart rate recordings during sleep where different segments
corresponding to different sleep stages exhibit different types of
correlations\cite{bundesleep2000}; ({\it ii}) DNA sequences including coding and
non-coding regions characterized by different correlations\cite{cknature1992,SMDFA1,HE1} 
and ({\it iii}) brain wave signals during different sleep stages\cite{SLEEP1}.

To better understand the complex behavior of $F(n)/n$ for such nonstationary
signals, we study their components separately. Each component is composed
only of those segments in the original signal which are  characterized by
identical correlations, 
while the segments with different correlations are substituted with zeros [see
Figs.~\ref{mix}(b) and (c)]. Since the two components of the nonstationary
signal in Fig.~\ref{mix}(a) are independent, based on the superposition rule
[Eq.~(\ref{mixadd1})], we expect that the r.m.s. fluctuation function $F(n)/n$
will behave as $\sqrt{[F_1(n)/n]^2+[F_2(n)/n]^2}$, where $F_1(n)/n$ and
$F_2(n)/n$ are the r.m.s. fluctuation functions of the components in
Fig.~\ref{mix}(b) and Fig.~\ref{mix}(c), respectively. We find a remarkable 
agreement between the superposition rule prediction and the result of the DFA
method obtained directly from the mixed signal [Fig~\ref{mix}(d)]. This
finding helps us understand the relation between the scaling behavior of
the mixed nonstationary signal and its components. 

Information on the effect 
of such parameters as the scaling exponents $\alpha_{1}$ and $\alpha_{2}$,
the size of the segments $W$ and their relative fraction $p$ on the scaling
behavior of the components provides insight into the scaling
behavior of the 
original mixed signal. When the original signal comes from real data,
its composition is {\it a priori} unknown. A first step is to ``guess'' the
type of correlations (exponents $\alpha_{1}$ and $\alpha_{2}$) present in the
signal, based on the scaling behavior of $F(n)/n$ at small and large scales
[Fig~\ref{mix}(d)]. A second step is to determine the parameters $W$ and $p$
for each component by matching the scaling result from the superposition rule
with the original signal. Hence in the following subsections, we focus on the
scaling properties of the components and how they change with $p$, $\alpha$
and $W$.

\subsubsection{Dependence on the fraction of segments}\label{fraction}

First, we study how the correlation properties of the components depend on
the fraction $p$ of the segments with identical local correlations. 

\begin{figure}
\centerline{
\epsfysize=0.9\columnwidth{\rotate[r]{\epsfbox{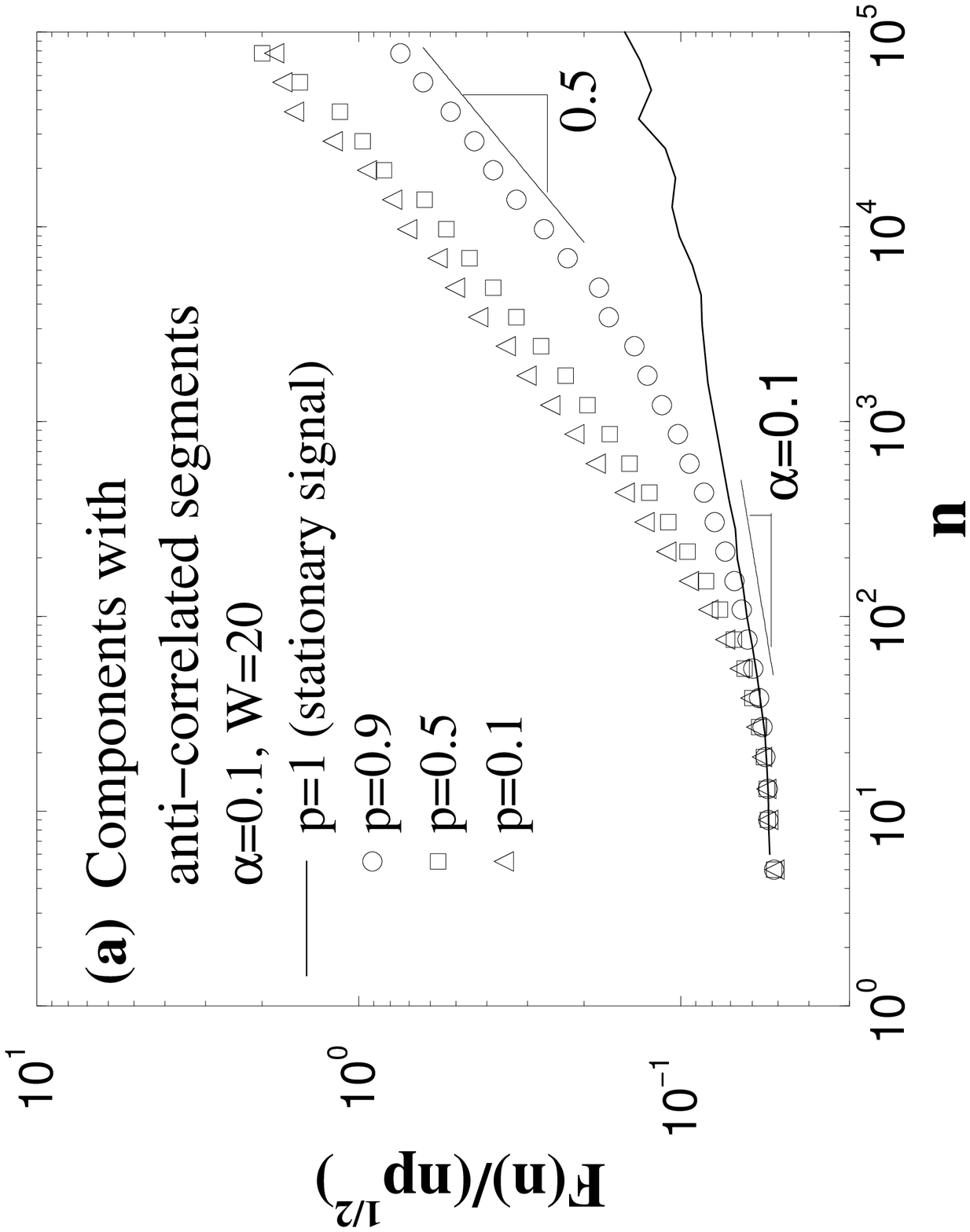}}}}
\centerline{
\epsfysize=0.9\columnwidth{\rotate[r]{\epsfbox{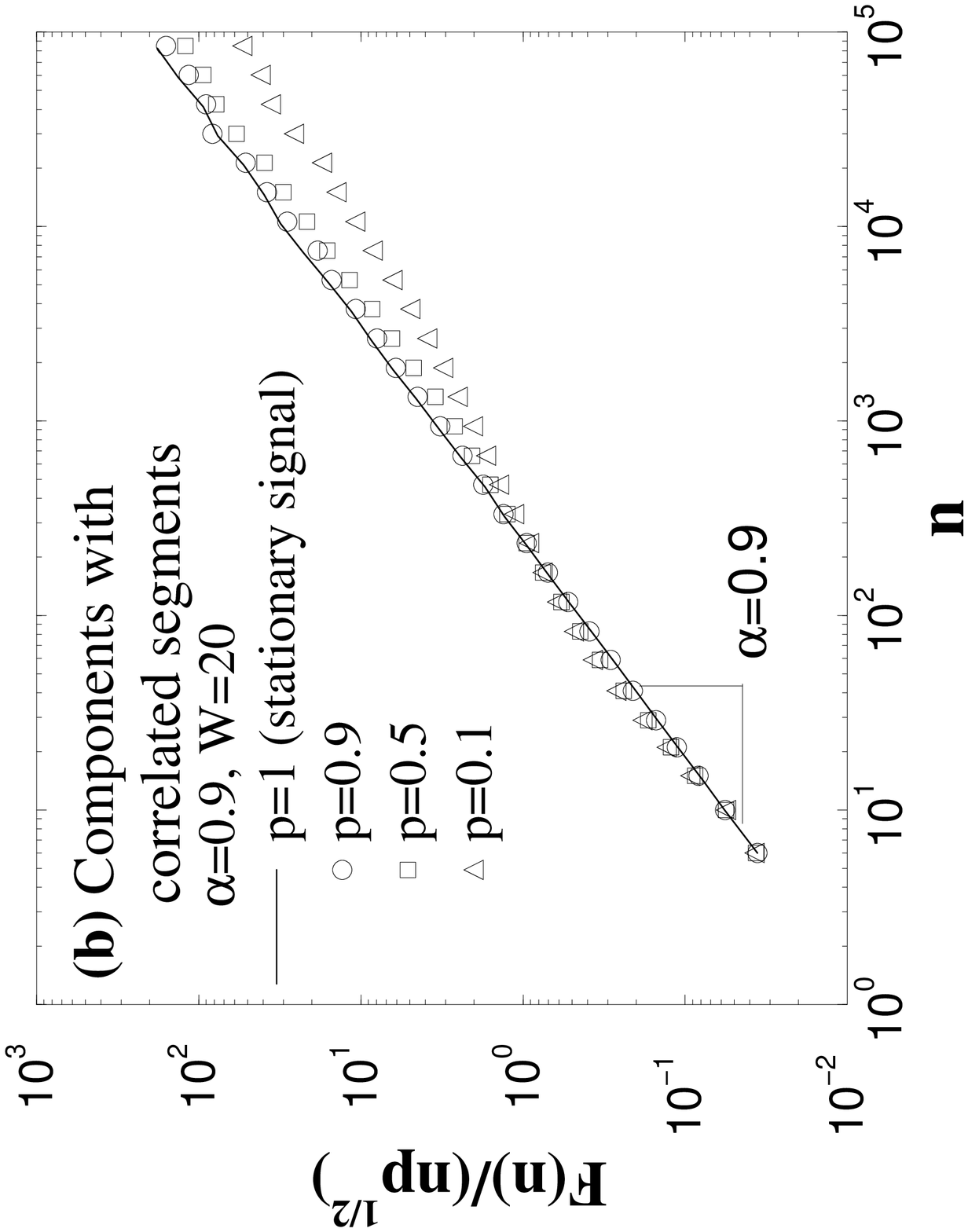}}}}
\caption{Dependence of the scaling behavior of components
  on the fraction $p$ of the segments with identical local correlations,
  emphasizing data collapse at {\it small\/} scales. The
  segment size is $W=20$ and the length of the components is
  $N_{max}=2^{20}$. (a) Components containing anti-correlated segments 
  ($\alpha=0.1$), at {\it small\/} scales ($n\leq W$). The slope of $F(n)/n$
  is identical to that expected for a stationary ($p=1$) signal with the same
  anti-correlations. After rescaling $F(n)/n$ by $\sqrt{p}$, at small scales
  all curves collapse on the curve for the stationary anti-correlated signal. 
 (b) Components containing correlated segments ($\alpha=0.9$), at {\it
  small\/} scales ($n\leq W$). The slope of $F(n)/n$ is identical to that
  expected for a 
  stationary ($p=1$) signal with the same correlations. After rescaling
  $F(n)/n$ by $\sqrt{p}$, at small scales all curves collapse on the curve
  for the stationary correlated signal.  
}
\label{mix2}
\end{figure}

For components containing segments with anti-correlations ($0<\alpha<0.5$)
and fixed size $W$ 
[Fig.~\ref{mix}(b)], we find a crossover to random behavior ($\alpha=0.5$) at
large scales, which becomes more pronounced (shift to smaller scales) when
 the fraction $p$ decreases [Fig.~\ref{mix2}(a)]. At {\it small \/} scales
($n\leq W$), the slope of 
$F(n)/n$ is identical to that expected for a stationary signal $u(i)$
(i.e., $p=1$) with the same anti-correlations [solid line in
Fig.~\ref{mix2}(a)]. Moreover, we observe a vertical shift in $F(n)/n$ to 
lower values when the fraction $p$ of non-zero anti-correlated segments
decreases. We find that at small scales after rescaling $F(n)/n$ by
$\sqrt{p}$, all curves collapse on the curve for the stationary
anti-correlated signal $u(i)$ [Fig.~\ref{mix2}(a)]. Since at small scales
($n\leq W$) the behavior of $F(n)/n$ does not depend on the segment size $W$,
this collapse indicates that the vertical shift in $F(n)/n$ is due only to
the fraction $p$. Thus, to determine the fraction $p$ of anti-correlated
segments in a nonstationary signal [mixture of anti-correlated and correlated
segments, Fig.~\ref{mix}(a)] we only need to estimate at small scales the
vertical shift in $F(n)/n$ 
between the mixed signal [Fig.~\ref{mix}(d)] and a stationary signal $u(i)$
with identical anti-correlations. This approach is valid for  
nonstationary signals where the fraction $p$ of the anti-correlated
segments is much larger than the fraction of the correlated segments in the
mixed signal [Fig.~\ref{mix}(a)], since only under this condition the
anti-correlated segments can dominate $F(n)/n$ of the mixed signal at small
scales [Fig.~\ref{mix}(d)].  

For components containing segments with positive correlations
 ($0.5<\alpha<1$) and fixed size $W$ 
 [Fig.~\ref{mix}(c)], we observe a similar behavior for $F(n)/n$, with collapse
 at {\it small} scales ($n\leq W$) after rescaling by $\sqrt{p}$
 [Fig.~\ref{mix2}(b)] (For $\alpha>1$, there are 
 exceptions with different rescaling factors, see
 Appendix~\ref{secadd2}). At small scales the values of $F(n)/n$ for
 components containing segments with positive correlations are  
 much larger compared to the values of $F(n)/n$ for components containing an
 identical fraction $p$ of anti-correlated segments
 [Fig.~\ref{mix2}(a)]. Thus, for a mixed signal where the fraction of
 correlated segments is not too small (e.g., $p\geq 0.2$), the contribution at
 small scales of the anti-correlated segments to $F(n)/n$ of the
 mixed signal [Fig.~\ref{mix}(d)] may not be observed, and the behavior (values
 and slope) of $F(n)/n$ will be dominated by the correlated segments. In this
 case, we must consider the behavior of $F(n)/n$ of the mixed signal at
 large scales only, since the contribution of the anti-correlated segments at
 large scales is negligible. Hence, we next study the scaling behavior of
 components with correlated segments at {\it large} scales. 

\begin{figure}
\centerline{
\epsfysize=0.9\columnwidth{\rotate[r]{\epsfbox{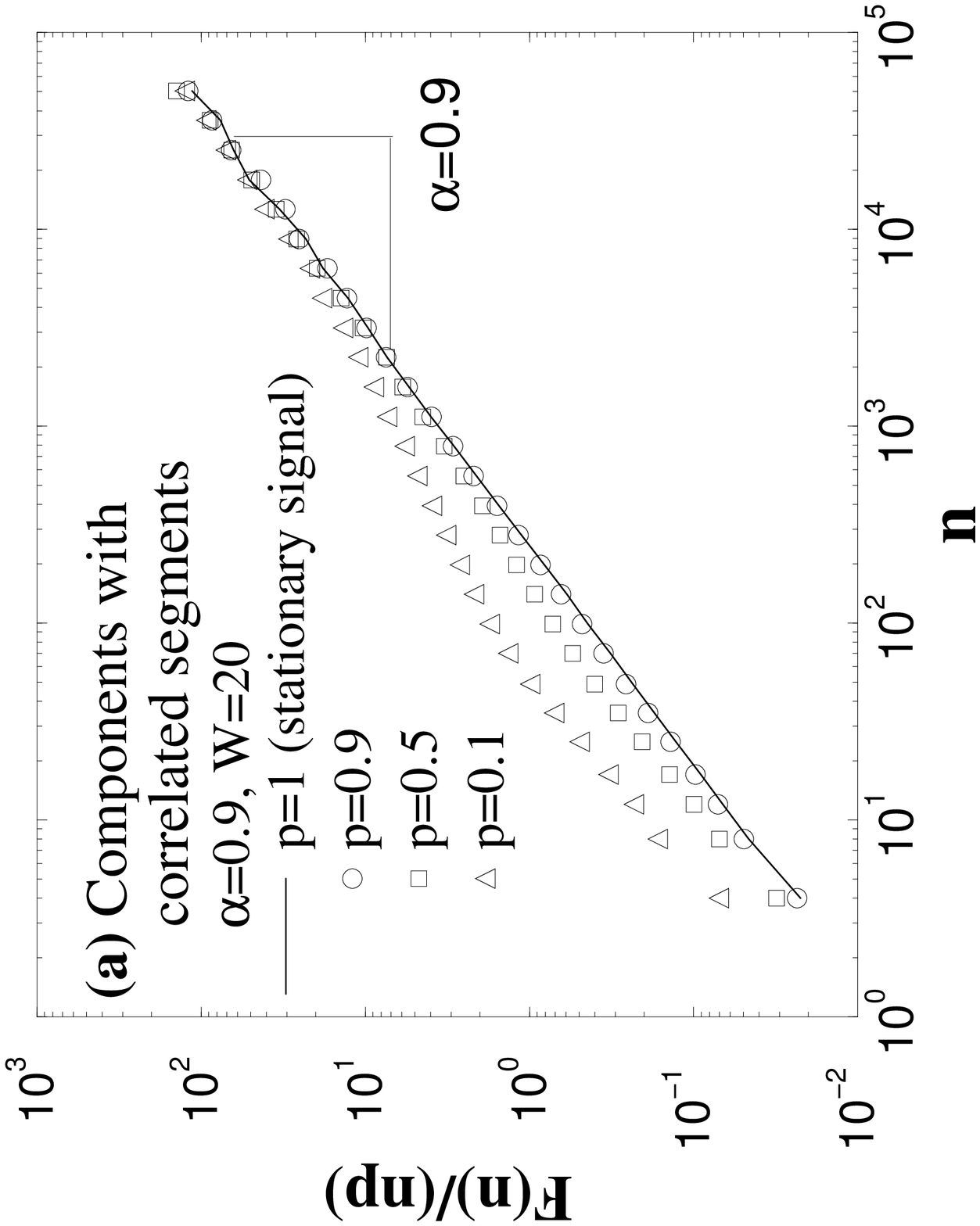}}}}
\centerline{
\epsfysize=0.9\columnwidth{\rotate[r]{\epsfbox{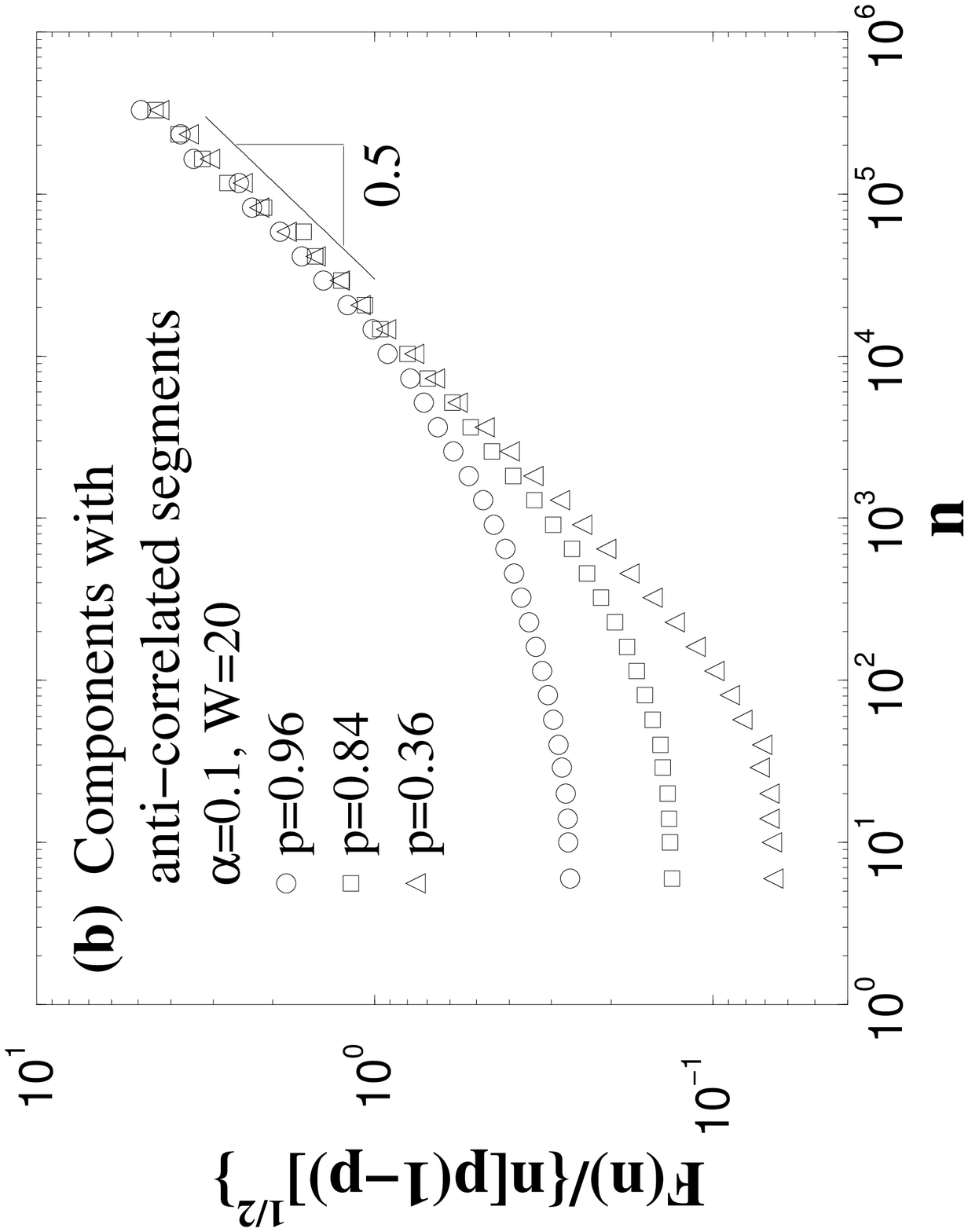}}}}
\vspace*{0.25cm}
\caption{Dependence of scaling behavior of components on the fraction $p$ of
  the segments with identical local correlations, emphasizing data collapse
  at {\it large\/} scales. The 
  segment size is $W=20$ and the length of the components is $N_{max}=2^{20}$. 
  (a) Components containing correlated segments ($\alpha=0.9$), at {\it
  large\/} scales ($n\gg W$). The slope of $F(n)/n$ is identical to that
  expected for a stationary ($p=1$) signal with the same correlations. After
  rescaling $F(n)/n$ by $p$, at large scales all curves collapse on the curve
  for the stationary correlated signal. (b) Components containing anti-correlated
  segments ($\alpha=0.1$), at {\it large\/} scales ($n\gg W$). There is a
  crossover to random behavior ($\alpha=0.5$).  After rescaling $F(n)/n$ by
  $\sqrt{p(1-p)}$, all curves collapse at large scales.  }
\label{mix3}
\end{figure}

For components containing segments with positive correlations
and fixed size $W$  [Fig.~\ref{mix}(c)], we find that at {\it large \/} scales
the slope of $F(n)/n$ is identical to that expected for a stationary
signal $u(i)$ (i.e., $p=1$) with the same correlations [solid line in
Fig.~\ref{mix3}(a)]. We also observe a vertical shift in $F(n)/n$ to lower
values when the fraction $p$ of non-zero correlated segments in the
component decreases. We find that after rescaling $F(n)/n$ by $p$, at large
scales all curves collapse on the curve representing the stationary
correlated signal $u(i)$ 
[Fig.~\ref{mix3}(a)]. Since at large scales ($n\gg W$), the effect of the zero
segments of size $W$ on the r.m.s. fluctuation function $F(n)/n$ for
components with correlated segments is negligible, even when the zero
segments are 50\% of the component [see Fig.~\ref{mix3}(a)], the finding of a collapse
at large scales indicates that the vertical shift in $F(n)/n$ is only due to
the fraction $p$ of the correlated segments. Thus, to determine the fraction
$p$ of correlated segments in a nonstationary signal (which is a mixture of
anti-correlated and correlated segments [Fig.~\ref{mix}(a)]), we only need to
estimate at large scales the vertical shift in $F(n)/n$ between the mixed signal
[Fig.~\ref{mix}(d)] and a stationary signal $u(i)$ with identical
correlations. 

For components containing segments with anti-correlations and fixed size $W$
[Fig.~\ref{mix}(b)], we find that at large scales in order to collapse the
$F(n)/n$ curves ($n\gg W$) [Fig.~\ref{mix2}(a)] we need to rescale $F(n)/n$ 
by  $\sqrt{p(1-p)}$ [see Fig.~\ref{mix3}(b)]. Note that there is a difference
between the rescaling factors for components with anti-correlated and
correlated segments at small [Figs.~\ref{mix2}(a-b)] and large  
[Figs.~\ref{mix3}(a-b)] scales. We also note that for components with
correlated segments ($\alpha>0.5$) and sufficiently small $p$, there is a
different rescaling factor of $\sqrt{p(1-p)}$ in the intermediate scale
regime [see Appendix~\ref{secadd2}, Fig.~\ref{mix6}]. 

For components containing segments of white noise ($\alpha=0.5$), we find no
change in the scaling exponent as a function of the fraction $p$ of the
segments, i.e., $\alpha=0.5$ for the components at both small and large
scales. However, we observe at all scales a vertical shift in $F(n)/n$ to
lower values with decreasing $p$: $F(n)/n \sim \sqrt{p}$.  

\subsubsection{Dependence on the size of segments}\label{width}

Next, we study how the scaling behavior of the components depends on the size
of the segments $W$. 

\begin{figure}
\centerline{
\epsfysize=0.9\columnwidth{\rotate[r]{\epsfbox{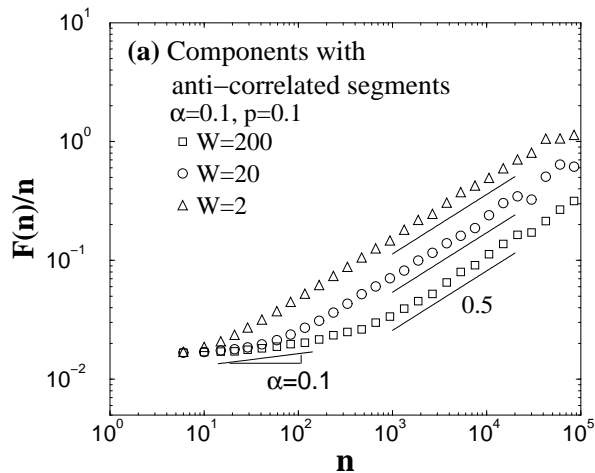}}}}
\centerline{
\epsfysize=0.9\columnwidth{\rotate[r]{\epsfbox{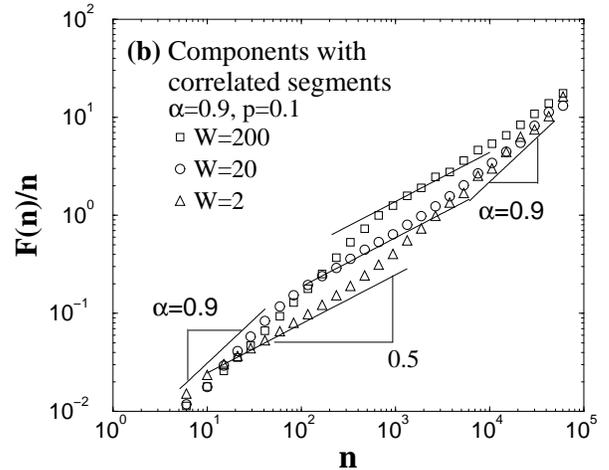}}}}
\vspace*{0.25cm}
\caption{Dependence of the scaling behavior of components on the segment size
  $W$. The fraction $p=0.1$ of the non-zero segments is
  fixed and the length of the components is $N_{max}=2^{20}$. (a) Components
  containing anti-correlated segments ($\alpha=0.1$). At large scales  
  ($n\gg W$), there is a crossover to random behavior ($\alpha=0.5$). 
   An equidistant vertical shift in $F(n)/n$ when $W$ decreases by a
  factor of ten suggests a power-law relation between
  $F(n)/n$ and $W$. (b) Components containing correlated segments
  ($\alpha=0.9$). At intermediate 
  scales, $F(n)/n$ has slope $\alpha=0.5$, indicating random behavior. An
  equidistant vertical shift in $F(n)/n$ suggests a power-law
  relation between $F(n)/n$ and $W$.  }
\label{mix4}
\end{figure}

First, we consider components containing segments with
anti-correlations. For a fixed value of the fraction $p$ of the segments, we 
study how $F(n)/n$ changes with $W$. At small scales, we observe a behavior
with a slope similar to the one for a stationary signal $u(i)$ with identical
anti-correlations [Fig.~\ref{mix4}(a)]. At large scales, we observe a
crossover to random behavior (exponent $\alpha=0.5$) with an increasing
crossover scale when $W$ increases. At large scales,   
we also find a vertical shift with increasing values for $F(n)/n$ when $W$
 decreases [Fig.~\ref{mix4}(a)]. Moreover, we find that there is an 
 equidistant vertical shift in $F(n)/n$ when $W$ decreases by a factor of
 ten, suggesting a power-law relation between
 $F(n)/n$ and $W$ at large scales.  

For components containing correlated segments with a fixed value of the
fraction $p$ we find that in the intermediate scale regime, the segment size $W$
plays an important role in the scaling behavior of $F(n)/n$  
[Fig.~\ref{mix4}(b)]. We first focus on the intermediate scale regime when
both $p=0.1$ and $W=20$ are fixed [middle curve in
Fig.~\ref{mix4}(b)]. We find that for a small fraction $p$ of the correlated
segments, $F(n)/n$ has slope $\alpha=0.5$, indicating random behavior
[Fig.~\ref{mix4}(b)] which shrinks when $p$ increases [see
Appendix~\ref{secadd2}, Fig.~\ref{mix6}]. Thus, for components containing  
 correlated segments, $F(n)/n$ approximates at large and small scales the
 behavior of a stationary signal with identical correlations ($\alpha=0.9$),
 while in the intermediate scale regime there is a plateau of random 
 behavior due to the random ``jumps'' at the borders between the non-zero and
 zero segments [Fig.~\ref{mix}(c)]. Next, we consider the case when the
 fraction of correlated segments $p$ is fixed while the segment 
 size $W$ changes. We find a vertical shift with increasing values for
 $F(n)/n$ when $W$ increases [Fig.~\ref{mix4}(b)], opposite to what we
 observe for components with anti-correlated segments
 [Fig.~\ref{mix4}(a)]. Since the vertical shift in $F(n)/n$ is equidistant
 when $W$ increases by a factor of ten, our finding indicates a power-law
 relationship between $F(n)/n$ and $W$.

\subsubsection{Scaling Expressions}\label{expression}

To better understand the complexity in the scaling behavior of components
with correlated and anti-correlated segments at different scales, we employ
the superposition rule (see \cite{kun} and Appendix~\ref{secadd}). For each
component we have
\begin{equation} 
F(n)/n=\sqrt{[F_{\rm corr}(n)/n]^2+[F_{\rm rand}(n)/n]^2},
\label{mixeq1}
\end{equation}
where $F_{\rm corr}(n)/n$ accounts for the contribution of the correlated or
anti-correlated non-zero segments, and $F_{\rm rand}(n)/n$ accounts for the
randomness due to ``jumps'' at the borders between non-zero and zero segments
in the component.
\\
\\
{\it Components with correlated segments ($\alpha>0.5$)} \\
\\
At small scales $n<W$, our findings presented in Fig.~\ref{mix2}(b)
suggest that there is no substantial contribution from $F_{\rm rand}(n)/n$. Thus 
 from Eq. (\ref{mixeq1}),  
\begin{equation} 
F(n)/n\approx F_{\rm corr}(n)/n\sim b_0\sqrt{p}n^{\alpha},
\label{mixeq2}
\end{equation}
where $b_0n^{\alpha}$ is the r.m.s. fluctuation function for stationary
($p=1$) correlated signals [Eq.~(\ref{si_bl}) and \cite{kun}]. 

\begin{figure}
\centerline{
\epsfysize=0.95\columnwidth{\rotate[r]{\epsfbox{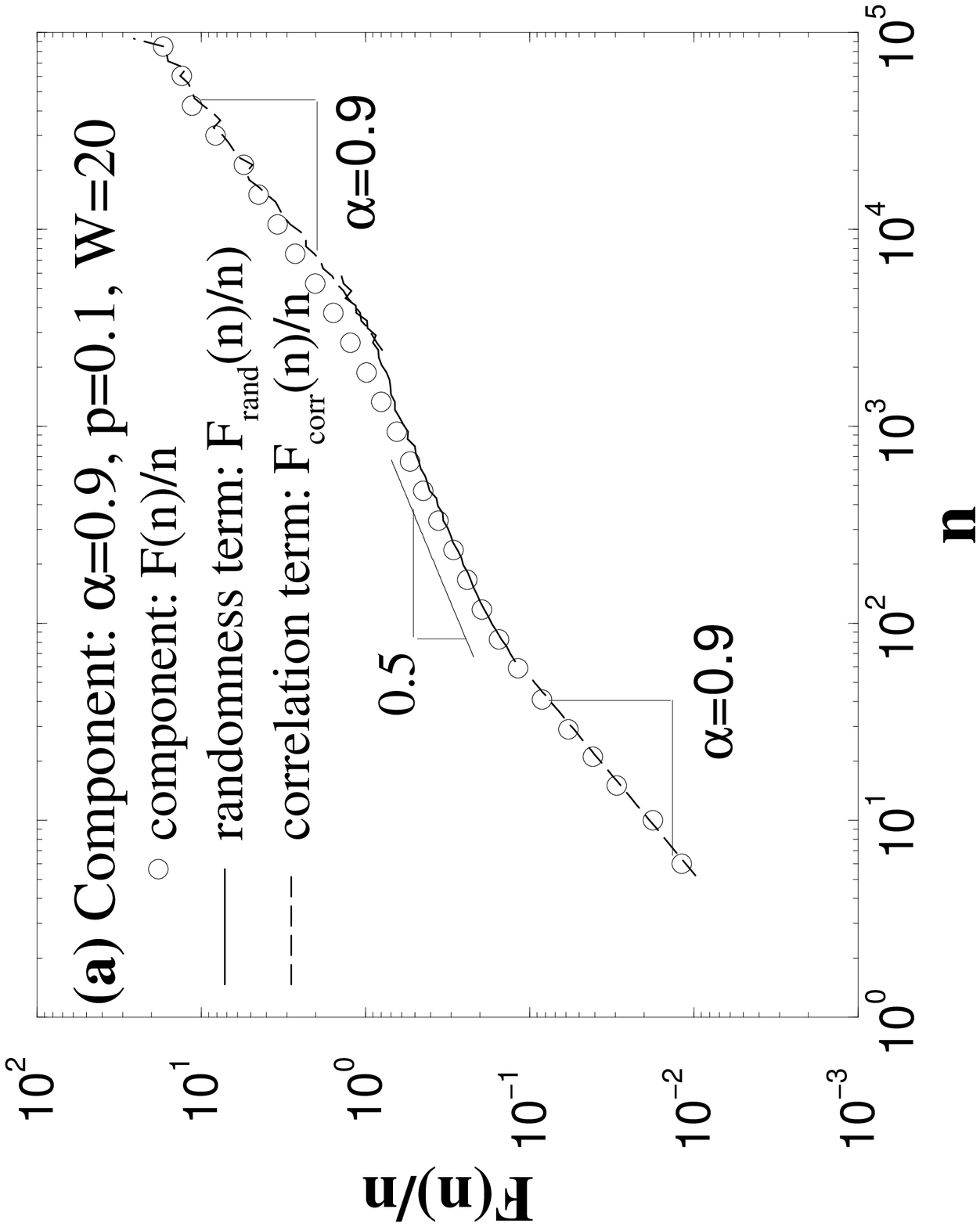}}}}
\centerline{
\epsfysize=0.9\columnwidth{\rotate[r]{\epsfbox{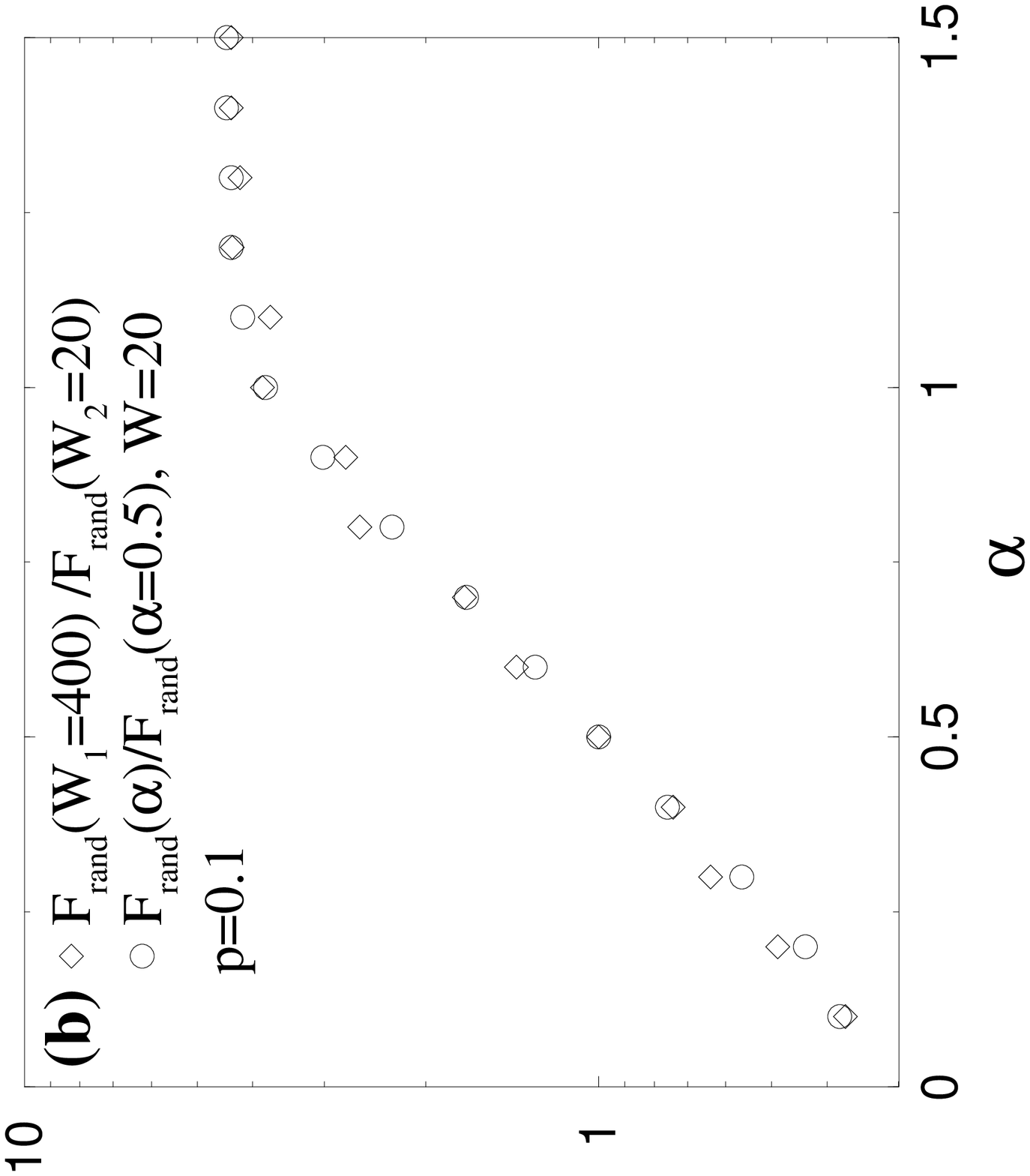}}}}
\vspace*{0.25cm}
\caption{ (a) Scaling behavior of components containing correlated segments
  ($\alpha>0.5$). $F(n)/n$ exhibits two crossovers and three scaling regimes
  at small, intermediate and large scales. From the superposition rule
  [Eq.~(\ref{mixeq1})] we find that the small and large scale regimes are
  controlled by the correlations ($\alpha=0.9$) in the segments
  [$F_{\rm corr}(n)/n$ from Eqs.~(\ref{mixeq2}) and~(\ref{mixeq3})] while the
  intermediate regime [$F_{\rm rand}(n)/n\sim n^{0.5}$ from Eq.~(\ref{mixeq4})]
  is dominated by the random jumps at the borders between non-zero and zero
  segments. (b) The ratio $F_{\rm rand}(W_1=400)/F_{\rm rand}(W_2=20)$ in the
intermediate scale regime for fixed $p$ and different values of $\alpha$, and
   the ratio $F_{\rm rand}(\alpha)/F_{\rm rand}(\alpha=0.5)$ for fixed $p$
  and $W=W_1/W_2$. $F_{\rm rand}(n)/n$ is obtained from Eq.~(\ref{mixeq4})
  and the ratios are estimated for all scales $n$ in the intermediate regime. 
 The two curves overlap for a broad range of values for the exponent $\alpha$,
  suggesting that $F_{\rm rand}(n)/n$ does not depend on $h(\alpha)$ [see
  Eqs.~(\ref{mixeq5}) and~(\ref{mixeq9})]. }
\label{mix5}
\end{figure}

Similarly, at large scales $n\gg W$, we find that the contribution of
$F_{\rm rand}(n)/n$ is negligible [see Fig.~\ref{mix3}(a)], thus from
Eq. (\ref{mixeq1}) we have
\begin{equation} 
F(n)/n\approx F_{\rm corr}(n)/n \sim b_0pn^{\alpha}.
\label{mixeq3}
\end{equation}
However, in the intermediate scale regime, the contribution of
$F_{\rm rand}(n)/n$ to $F(n)/n$ is substantial. To confirm this we use the
superposition rule 
[Eq.~(\ref{mixeq1})] and our estimates for $F_{\rm corr}(n)/n$ at small
[Eq.~(\ref{mixeq2})] and large [Eq.~(\ref{mixeq3})] scales\cite{note2}. The result we
obtain from
\begin{equation} 
F_{\rm rand}(n)/n=\sqrt{[F(n)/n]^2-[b_0\sqrt{p}n^{\alpha}]^2-[b_0pn^{\alpha}]^2}
\label{mixeq4}
\end{equation}   
 overlaps with $F(n)/n$ in the intermediate scale regime, exhibiting  a
slope of $\approx 0.5$: $F_{\rm rand}(n)/n\sim n^{0.5}$
[Fig.~\ref{mix5}(a)]. Thus, $F_{\rm rand}(n)/n$ is indeed a contribution due to
the random jumps between the non-zero correlated segments and the zero
segments in the component [see Fig.~\ref{mix}(c)]. 

Further, our results in Fig.~\ref{mix4}(b) suggest that in the intermediate
scale regime $F(n)/n\sim W^{g_c(\alpha)}$ for fixed fraction $p$ [see
Sec.~\ref{width}], where the
power-law exponent $g_c(\alpha)$ may be a 
function of the scaling exponent $\alpha$ characterizing the correlations in
the non-zero segments. Since at intermediate scales $F_{\rm rand}(n)/n$ dominates
 the scaling [Eq.~(\ref{mixeq4}) and Fig.~\ref{mix5}(a)], from
 Eq.~(\ref{mixeq1}) we find $F_{\rm rand}(n)/n \approx F(n)/n
\sim W^{g_c(\alpha)}$. We also find that at intermediate scales, $F(n)/n\sim
\sqrt{p(1-p)}$ for fixed segment size $W$ (see Appendix~\ref{secadd2},
Fig.~\ref{mix6}). Thus from Eq.~(\ref{mixeq1}) we find $F_{\rm rand}(n)/n \approx
F(n)/n \sim \sqrt{p(1-p)}$. Hence we obtain the following general expression
\begin{equation}
F_{\rm rand}(n)/n\sim h(\alpha)\sqrt{p(1-p)}W^{g_c(\alpha)}n^{0.5}.
\label{mixeq5}
\end{equation}
Here we assume that $F_{\rm rand}(n)/n$ also depends directly on the type of
correlations in the segments through some function $h(\alpha)$.

To determine the form of $g_c(\alpha)$ in Eq.~(\ref{mixeq5}), we perform the
following steps:\\
(a) We fix the values of $p$ and $\alpha$, and from Eq.~(\ref{mixeq4}) we
calculate the value of $F_{\rm rand}(n)/n$ for two different values of the
segment size $W$, e.g., we choose $W_1=400$ and $W_2=20$.\\
(b) From the expression in Eq.~(\ref{mixeq5}), at the same scale $n$ in the
intermediate scale regime we determine the ratio:
\begin{equation}
F_{\rm rand}(W_1)/F_{\rm rand}(W_2)=(W_1/W_2)^{g_c(\alpha)}.
\label{mixeq6}
\end{equation} 
(c) We plot $F_{\rm rand}(W_1)/F_{\rm rand}(W_2)$ vs. $\alpha$ on a linear-log scale
[Fig.~\ref{mix5}(b)]. From the graph and Eq.~(\ref{mixeq6}) we obtain the
dependence  
\begin{eqnarray}
g_c(\alpha)&=&\frac{\log[F_{\rm rand}(W_1)/F_{\rm
    rand}(W_2)]}{\log(W_1/W_2)}\nonumber\\
&=&\left\{\begin{array}{l}
C\alpha-C/2,\mbox{ $0.5\leq \alpha \leq 1$}\\
0.50, \mbox{ for $\alpha>1$},
\end{array} \right.
\label{mixeq7}
\end{eqnarray}
where $C=0.87\pm0.06$. Note that $g_c(0.5)=0$.

To determine if $F_{\rm rand}(n)/n$ depends on $h(\alpha)$ in Eq.~(\ref{mixeq5}), we
perform the following steps:\\
(a) We fix the values of $p$ and $W$ and 
calculate the value of $F_{\rm rand}(n)/n$ for two different values of the
scaling exponent $\alpha$, e.g., $0.5$ and any other value of $\alpha$ from
Eq.~(\ref{mixeq4}).\\ 
(b) From the expression in Eq.~(\ref{mixeq5}), at the same scale $n$ in 
the intermediate scale regime we determine the ratio:
\begin{eqnarray}
\frac{F_{\rm rand}(\alpha)}{F_{\rm rand}(0.5)}=\frac{h(\alpha)}{h(0.5)}
W^{g_c(\alpha)-g_c(0.5)}= \frac{h(\alpha)}{h(0.5)} W^{g_c(\alpha)},
\label{mixeq8}
\end{eqnarray} 
since $g_c(0.5)=0$ from Eq.~(\ref{mixeq7}).\\
(c) We plot $F_{\rm rand}(\alpha)/F_{\rm rand}(0.5)$ vs.~$\alpha$ on a linear-log 
scale [Fig.~\ref{mix5}(b)] and find that when $W\equiv W_1/W_2$ [in
Eqs.~(\ref{mixeq6}) and~(\ref{mixeq8})] this curve 
overlaps with $F_{\rm rand}(W_1)/F_{\rm rand}(W_2)$ vs.~$\alpha$
[Fig.~\ref{mix5}(b)] for all values of the scaling exponent 
$0.5\leq \alpha\leq 1.5$. From this overlap and from Eqs.~(\ref{mixeq6}) 
and~(\ref{mixeq8}), we obtain  
\begin{equation}
W^{g_c(\alpha)}=\frac{h(\alpha)}{h(0.5)}W^{g_c(\alpha)}
\label{mixeq8b}
\end{equation}
for every value of $\alpha$, suggesting that
  $h(\alpha)=const$ and thus $F_{\rm rand}(n)/n$ can finally be expressed as:
\begin{equation}
F_{\rm rand}(n)/n\sim \sqrt{p(1-p)}W^{g_c(\alpha)}n^{0.5}.
\label{mixeq9}
\end{equation}
\\
{\it Components with anti-correlated segments ($\alpha<0.5$)} \\
\\
Our results in Fig.~\ref{mix2}(a) suggest that at small
scales $n<W$ there is no substantial contribution of $F_{\rm rand}(n)/n$ and
that: 
\begin{equation} 
F(n)/n\approx F_{\rm corr}(n)/n \sim b_0\sqrt{p}n^{\alpha},
\label{mixeq10}
\end{equation}
a behavior similar to the one we find for components with correlated segments
[Eq.~(\ref{mixeq2})]. 

In the intermediate and large scale regimes ($n\geq W$), from the plots in
Fig.~\ref{mix3}(b) and Fig.~\ref{mix4}(a) we find the scaling behavior of
$F(n)/n$ is controlled by $F_{\rm rand}(n)/n$ and thus
\begin{equation}
F(n)/n\approx F_{\rm rand}(n)/n\sim \sqrt{p(1-p)}W^{g_a(\alpha)}n^{0.5}, 
\label{mixeq11}
\end{equation}
where $g_a(\alpha)=C\alpha-C/2$ for $0<\alpha<0.5$ [see Fig.~\ref{mix5}(b)] and
the relation for $F_{\rm rand}(n)/n$ is obtained using the same procedure we
followed for Eq.~(\ref{mixeq9}).

\section{Conclusions}\label{Conclusion}

In this paper we studied the effects of three different types of
nonstationarities using the DFA correlation analysis method.
Specifically, we consider sequences formed in three ways: ({\it i})
stitching together segments of signals obtained from
discontinuous experimental recordings, or removing some noisy and unreliable
 segments from continuous recordings and stitching together the remaining
 parts; ({\it ii}) adding random outliers or spikes to a signal with known
 correlations, and ({\it iii}) generating a signal 
composed of segments with different properties --- e.g. different standard
deviations or different correlations. We 
compare the difference between the scaling results obtained for stationary
correlated signals and for correlated signals with artificially imposed 
 nonstationarities. 
\\

({\it i}) We find that removing segments from a signal and
 stitching together the remaining parts does not affect the scaling behavior
 of 
 positively correlated signals ($1.5 \geq \alpha > 0.5$), even when up to 50\%
 of the points in these signals are removed. However, such a    
cutting procedure strongly affects anti-correlated signals, leading to a 
crossover from an anti-correlated regime (at small scales) to an uncorrelated 
regime (at large scales). The crossover scale $n_{\times}$ increases with 
increasing value of the scaling exponent $\alpha$ for the original stationary 
anti-correlated signal. It also depends both on 
the segment size and the fraction of the points cut out from the signal: (1)  
$n_{\times}$ decreases with increasing fraction of cutout segments, and (2)
$n_{\times}$ increases with increasing segment size. Based on our findings,
we propose an approach to minimize the effect of cutting procedure on the
correlation properties of a signal: When two segments which need to be
removed are on distances shorter than the size of the segment, it is
advantageous to cut out both the segments and the part of the signal between
them.         
\\

({\it ii}) Signals with superposed random spikes. We find that for
an anti-correlated signal with superposed spikes at  
small scales, the scaling behavior is close to that of the stationary 
anti-correlated signal without spikes. At large scales, there is a crossover 
to random behavior. For a correlated signal with spikes, we find 
a different crossover from uncorrelated behavior at small scales to 
correlated behavior at large scales with an exponent close to the exponent of 
the original stationary signal. We also find that the spikes with identical
density and amplitude may cause strong effect on the scaling of an
anti-correlated signal while they 
have a much smaller or no effect on the scaling of a correlated signal --- when 
the two signals have the same standard deviations. We investigate the
characteristics of the scaling of the spikes only and find that 
the scaling behavior of the signal with random spikes is a superposition 
of the scaling of the signal and the scaling of the spikes. We 
analytically prove this superposition relation by introducing a 
{\it superposition rule\/}. 
\\ 

({\it iii}) Signals composed of segments with different local properties. We
find that  

(a) For nonstationary correlated signals comprised of segments which are
 characterized by two different values of the standard deviation, there is no
 difference in the scaling behavior compared to stationary correlated signals
 with constant standard deviation. For nonstationary anti-correlated signals,
 we find a crossover at scale $n_{\times}$. For small scales $n<n_{\times}$,
 the scaling behavior is similar to that of the stationary anti-correlated
 signals with constant standard deviation. For large scales $n>n_{\times}$,
 there is a transition to random behavior. We also find that very few
 segments with large standard deviation can strongly affect the
 anti-correlations in the signal. In contrast, the same fraction of segments
 with standard deviation smaller than the standard deviation of the rest of
 the anti-correlated signal has much weaker effect on the scaling behavior
 --- $n_{\times}$ is shifted to larger scales.

(b) For nonstationary signals consisting of segments with different 
correlations, the scaling behavior is a superposition of 
the scaling of the different components --- where 
each component contains only the segments exhibiting identical correlations 
and the remaining segments are replaced by zero. Based on our findings, we
propose an approach to identify the composition of such complex signals:
A first step is to ``guess'' the type of correlations from the DFA  
results at small and large scales. A second step is to determine the
parameters defining the components, 
such as the segment size and the fraction of non-zero segments. We studied 
how the scaling characteristics of the components depend on these parameters 
and provide analytic scaling expressions.

\section*{Acknowledgments} 
We thank NIH/National Center for Research Resources (Grant No.~P41RR13622)
and NSF for support. We also thank Pedro Carpena, C.-K. Peng, Jan
W. Kantelhardt and Verena Schulte-Frohlinde  for reading the manuscript and
for helpful suggestions.

\appendix

\section{Superposition rule}\label{secadd}

Here we show how the DFA results for any two signals $f$ and
 $g$ [denoted as $F_f(n)$ and $F_g(n)$] relate with the DFA result for the
 sum of these two signals $f+g$ [denoted as $F_{f+g}(n)$, where $n$ is the box
 length (scale of analysis)]. In the general cases, we find 
 $|F_f-F_g|\leq F_{f+g} \leq F_f+F_g$.  When the two signals are not 
correlated, we find that the following {\it superposition rule} is valid:
$F^2_{f+g}=F^2_{f}+F^2_{g}$. Here we derive these relations.

First we summarize again the procedure of the DFA method\cite{CKDFA1}. It
includes the following steps: starting with an original signal $u(i)$ of length
$N_{max}$, we integrate and
obtain $y(k)=\sum\limits_{j=1}^{k}(u(j)-\langle u\rangle)$, where
$\langle u\rangle$ is the mean of $u(i)$. Next, we divide $y(k)$ 
 into non-overlapping boxes of equal length $n$. In each box we
 fit the signal $y(k)$ using a polynomial function
$y_n(k)=a_0+a_1x(k)+a_2x^2(k)+...+a_sx^s(k)$, where $x(k)$ is the $x$
coordinate corresponding to the $k$th signal point. We calculate 
the r.m.s. fluctuation function 
$F(n)=\sqrt{\frac{1}{N_{max}}\sum\limits^{N_{max}}_{k=1} [y(k)-y_{n}(k)]^2}$. 

To prove the superposition rule, we first focus on one particular box along
the signal. In order to find the analytic expression of best fit in this box,
we write  
\begin{eqnarray}
I(a_0,...,a_s)
=\sum\limits_{k=1}^{n}[y(k)-(a_0+...+a_sx^s(k))]^2,
\label{eqn4}
\end{eqnarray}
where $a_m, m=0,...,s$ are the same for all points in 
 this box. ``Best fit'' requires that $a_m, m=0,...,s$ satisfy
\begin{eqnarray}
\frac{\partial I}{\partial a_m}=0,m=0,...s
\label{eqn5}
\end{eqnarray}
Combining Eq.~(\ref{eqn4}) with~(\ref{eqn5}) we obtain $s+1$ equations
\begin{eqnarray}
y_{m}=a_0t_{m0}+a_1t_{m1}...+a_st_{ms}, m=0,...,s
\label{eqn6}
\end{eqnarray}
where
\begin{eqnarray}
y_{m}=\sum\limits_{k=1}^{n} y(k) x^m(k), t_{mj}=\sum\limits_
{k=1}^{n} x^{m+j}(k), j=0,...,s
\label{eqn7}
 \end{eqnarray}

From Eqs.~(\ref{eqn6}) we determine $a_0,a_1,..,a_s$.

For the signals $f$, $g$ and $f+g$ after the integration, in each box we have  
\begin{eqnarray}
& &f_{m}=a_0t_{m0}+a_1t_{m1}...+a_st_{ms},m=0,...,s 
\nonumber \\
& &g_{m}=a_0^\prime t_{m0}+a_1^\prime t_{m1}...+a_s^\prime 
t_{ms},m=0,...,s \nonumber\\
& &(f+g)_{m}=a_0^{\prime \prime}t_{m0}+a_1^{\prime
\prime} t_{m1}...+a_s^{\prime \prime}
t_{ms},m=0,...,s 
\label{eqn09}
\end{eqnarray}
where $f_{m}$, $g_{m}$ and $(f+g)_{m}$ correspond to $y_{m}$ in Eqs.~(\ref{eqn6}).
  
Comparing the three groups of equations in Eqs.~(\ref{eqn09}), we find that,
when we add the first two groups together, the left side becomes
$f_{m}+g_{m}=(f+g)_{m}$ , which is precisely the left side of the third group of
equations. Thus we find 
\begin{eqnarray}
a_m^{\prime \prime}=a_m+a_m^\prime,m=0,...,s
\label{eqn12}
\end{eqnarray}
and for each point $k$ in every box, the polynomial fits for the signals
$f$, $g$ and $f+g$ satisfy
\begin{eqnarray}
(f+g)_{n}(k)=f_{n}(k)+g_{n}(k). 
\label{eqn13}
\end{eqnarray}
This result can be extended to all boxes in the signals. For the signal $f+g$
we obtain 
\begin{eqnarray}
F^2_{f+g}&=&\frac{1}{N_{max}}\sum\limits_{k=1}^{N_{max}}[f(k)-
f_{n}(k)]^2+[g(k)-g_{n}(k)]^2 \nonumber \\
& &+2[f(k)-f_{n}(k)][g(k)-g_{n}(k)].
\label{eqn17}
\end{eqnarray}
After the substitutions $f(k)-f_{n}(k)=Y_f(k)$ and 
$g(k)-g_{n}(k)=Y_g(k)$, we rewrite the above equation as
\begin{eqnarray}
 F^2_{f+g}&=&\frac{1}{N_{max}}\Bigl [\sum\limits_{k=1}^{N_{max}}(Y_f(k))^2+\sum\limits_{k=1}^{N_{max}}(Y_g(k))^2\nonumber \\
& &+2\sum\limits_{k=1}^{N_{max}}Y_{f}(k)Y_{g}(k)\Bigr ]\nonumber \\
&=&F^2_{f}+F^2_{g}+\frac{2}{N_{max}}\sum\limits_{k=1}^{N_{max}}Y_{f}(k)Y_{g}(k).
\label{eqn18}
\end{eqnarray}

In the general case, we can utilize the Cauchy inequality
\begin{eqnarray}
& &\left|\sum\limits_{k=1}^{N_{max}}Y_f(k)Y_g(k)\right|\nonumber  \\
& &\leq\left(\sum\limits_{k=1}^{N_{max}}(Y_f(k))^2\right)^{1/2}\left(
\sum\limits_{k=1}^{N_{max}}
(Y_g(k))^2\right)^{1/2}
\label{eqn19}
\end{eqnarray}
and we find 
\begin{eqnarray}
&&(F_f-F_g)^2 \leq F^2_{f+g} \leq (F_f+F_g)^2\nonumber \\
&\Longrightarrow &|F_f-F_g|\leq F_{f+g} \leq F_f+F_g.
 \label{eqn20}
\end{eqnarray}

From Eqs.~(\ref{eqn6}) for $m=0$, in every box we have $\sum\limits_{k=1}^{n} 
y(k)=\sum\limits_{k=1}^{n} y_{n}(k)$. Thus we obtain
$\sum\limits_{k=1}^{N_{max}}Y_f(k)=\sum\limits_{k=1}^{N_{max}}Y_g(k)=0$ where
$Y_f(k)$ and $Y_g(k)$ fluctuate around zero. 
When $Y_f(k)$ and $Y_g(k)$ are not correlated, the value of the third
term in Eq.~(\ref{eqn18}) is close to zero and we obtain the following
superposition rule  
\begin{eqnarray}
F^2_{f+g}=F^2_{f}+F^2_{g}.
\label{eqnadd1}
\end{eqnarray}

\section{Strongly correlated segments}\label{secadd2}

\begin{figure}
\centerline{
\epsfysize=0.9\columnwidth{\rotate[r]{\epsfbox{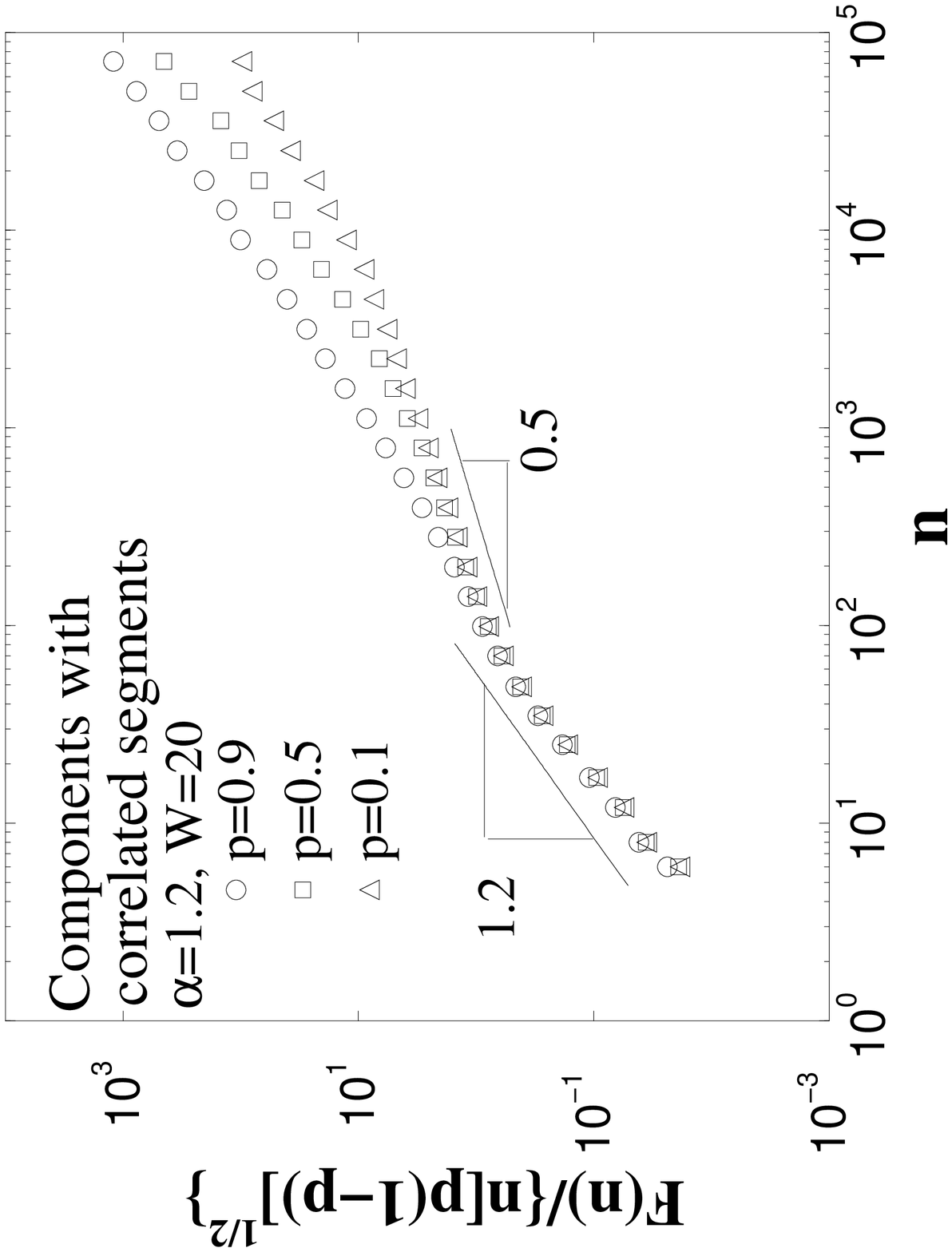}}}}
\vspace*{0.25cm}
\caption{Dependence of the scaling behavior of components on the fraction $p$
  of the segments with strong positive correlations ($\alpha=1.2$). The
  segment size is $W=20$ and the length of the components is
  $N_{max}=2^{20}$. After rescaling $F(n)/n$ by $\sqrt{p(1-p)}$, all curves
  collapse at small scales ($n<W$) with slope $1.2$ and at intermediate scales
  with slope $0.5$. The intermediate scale regime shrinks when $p$ increases.} 
\label{mix6}
\end{figure}

For components containing segments with stronger positive correlations
($\alpha>1$) and fixed $W=20$, we find that at small scales ($n<W$), the
slope of $F(n)/n$ does not depend on the fraction $p$ and is close to that
expected for a stationary signal $u(i)$ with identical correlations
(Fig.~\ref{mix6}). Surprisingly we find 
that in order to collapse the $F(n)/n$ curves obtained for different values
of the fraction $p$, we need to rescale $F(n)/n$ by $\sqrt{p(1-p)}$ instead
of $\sqrt{p}$, which is the rescaling factor at small scales for components 
containing segments with correlations $\alpha<1$. Thus  
$\alpha=1$ is a threshold indicating when the rescaling factor
changes. Our simulations show that this threshold increases when the
segment size $W$ increases.      

For components containing a sufficiently small fraction $p$ of correlated
segments ($\alpha>0.5$), we find that in the intermediate scale regime there is a
crossover to random behavior with slope $0.5$. The $F(n)/n$ curves obtained
for different values of $p$ collapse in the intermediate scale regime if we
rescale $F(n)/n$ by $\sqrt{p(1-p)}$ (Fig.~\ref{mix6}). We note that this
random behavior regime at intermediate scales shrinks with increasing the
fraction $p$ of correlated segments and, as expected, for $p$ close to $1$
this regime disappears (see the $p=0.9$ curve in Fig.~\ref{mix6}).   

\end{multicols}


\begin{thebibliography}{99}
\bibitem{gamma} When $0<\gamma<1$, $\alpha$ and $\gamma$ have the following
  relation: $ \alpha=(2-\gamma)/2$.
\bibitem{note2} We note that, for components containing strongly correlated 
segments (e.g., $\alpha=1.2$ when $W=20$, see Appendix~\ref{secadd2} and 
Fig.~\ref{mix6}), at small scales the contribution of the correlated non-zero 
segments [$F_{\rm corr}(n)/n$] is still substantial, however, we have 
$F(n)/n\approx F_{\rm corr}(n)/n\sim A\sqrt{p(1-p)}n^{\alpha}$.
\end{thebibliography}
\end{document}